\documentclass[journal]{IEEEtran}
\usepackage{cite}
\usepackage{amsmath,amssymb,amsfonts}
\usepackage{hyperref}
\usepackage{algorithmic}
\usepackage{array}
\usepackage[caption=false]{subfig}
\usepackage{textcomp}
\usepackage{stfloats}
\usepackage{url}
\usepackage{verbatim}
\usepackage{graphicx}
\graphicspath{{figure/}{./}}
\def\BibTeX{{\rm B\kern-.05em{\sc i\kern-.025em b}\kern-.08em
    T\kern-.1667em\lower.7ex\hbox{E}\kern-.125emX}}
\usepackage{balance}

\begin{document}
\title{Geometry-Aware Multi-UAV Full-Duplex Communication: System Design and Experiment}
\author{Tao Yu,~\IEEEmembership{Member,~IEEE}, 
Kiyomichi Araki,
Tomohiro Mogi,
Yasushi Hada, 
Kei Sakaguchi,~\IEEEmembership{Senior~Member,~IEEE}

\thanks{Copyright (c) 2026 IEEE. Personal use of this material is permitted. However, permission to use this material for any other purposes must be obtained from the IEEE by sending a request to pubs-permissions@ieee.org.}
\thanks{T. Yu, K. Araki, and K. Sakaguchi are with Institute of Science Tokyo, Tokyo, Japan (e-mail: \{yutao, araki, sakaguchi\}@mobile.ee.titech.ac.jp).}
\thanks{T. Mogi is with Koden Electronics Co., Ltd., Japan (e-mail: t-mogi@koden-electronics.co.jp). }
\thanks{Y. Hada is with the Department of Mechanical Systems Engineering, Kogakuin University, Japan (e-mail: had@cc.kogakuin.ac.jp).}
\thanks{This work was partially supported by the Japanese Ministry of Internal Affairs and Communications (MIC) under the grant agreement 0155-0083.}}


\maketitle

\begin{abstract}
The deployment of unmanned aerial vehicle (UAV) systems relies on high-performance yet lightweight wireless links between UAVs and ground stations (GSs). This paper presents a geometry-aware multi-UAV in-band full-duplex (MU-IBFD) communication system that uses high-gain directional antennas and separated uplink/downlink channels to convert self-interference into controllable co-channel interference (CCI) between UAVs, thereby avoiding energy-intensive self-interference cancelers on UAVs. We also derive a geometry-aware CCI model and define a reliable operating region (ROR) in the 3D airspace, within which the SINR requirement is satisfied.
A prototype consisting of two UAVs and a GS is developed, and field trials are conducted. The measured CCI as a function of UAV positions agrees well with the theoretically predicted non-ROR region, and the downlink capacity significantly exceeds that of a conventional time-division duplex (TDD) with omni-directional scheme and higher transmit power and approaches that of ideal IBFD in most of the airspace. A proof-of-concept 4K/60p video transmission further demonstrates the practical potential of the proposed MU-IBFD system.
\end{abstract}


\section{Introduction}
Over the past decade, unmanned aerial vehicles (UAVs) have experienced transformative progress, evolving into more operable, functional, productive, and consumer-affordable systems. This evolution is largely attributed to the miniaturization of electronic components, and enhanced flight control algorithms. In contrast to terrestrial robots, UAVs features the unique capability to extend their mobility into three-dimensional (3D) space, thereby offering an unprecedented level of flexibility and adaptability. This unique attribute positions them as suitable mobile platforms for remote and long-distance applications. They hold immense potential for a wide range of applications, including industrial uses such as infrastructure inspections, geodetic surveys, disaster relief, and emergency communications \cite{1}, as well as fast-growing personal uses such as aerial photography, nature observation, and sports/outdoor activities recording. The market forecast for UAVs and their associated applications predicts that by 2030, their market value will reach an impressive 54.6 billion USD \cite{2}.

To fully exploit the 3D mobility of UAVs and the advancement of UAV-based applications, high-performance wireless aerial communication systems are essential. They facilitate real-time data transmission, and thus act as a vital link between UAVs and ground stations (GSs). For remote applications, the ideal aerial communication system must offer wide coverage and high throughput, while simultaneously supporting multiple aerial connections for numerous UAVs. However, considering the constraints of UAV platforms, such as limited battery life, payload capacity, and size, it is crucial for these communication systems to maintain low hardware complexity, low energy consumption, and low cost.
Achieving aerial communications that satisfy these requirements faces unique challenges. UAVs generally operate at altitudes lacking natural or man-made obstacles like buildings or hills, which makes wireless channels between UAVs nearly constant line-of-sight (LOS). Such characteristics result in significant interference when multiple UAVs are involved, severely degrading system performance and necessitating additional costs for implementing carrier sensing and collision avoidance mechanisms.
Consequently, the system is compelled to inefficiently reuse radio resources through, e.g., time-division (TD) and frequency-division (FD), which results in a contrast between the high requirements of aerial communication systems and the low performance of conventional multiple UAV aerial communication systems, particularly when considering the limited frequency resource allocated to UAVs by regulations and laws.

Currently, most commercial aerial communication systems rely on LTE or WiFi-based solutions, or their derivatives. The 3GPP has investigated the performance of LTE networks for aerial communications in \cite{4}, and the interference between aerial and terrestrial users has been empirically evaluated in \cite{5}. Numerous works have explored UAV communication systems utilizing IEEE 802.11a/b/n/ac/g/s standards, as summarized in \cite{3}. However, the cellular and IEEE 802.11 families were initially designed for terrestrial use. As a result, their system configurations, multiplexing/duplexing schemes, and interference avoidance mechanisms are not ideally suited for aerial communication. For example, the carrier sensing multiple access with collision avoidance (CSMA/CA) mechanism exhibits reduced efficiency in aerial environments due to lower interference attenuation and longer propagation delays compared to terrestrial environments. Additionally, both LTE and WiFi are either half-duplex or out-band full-duplex (OBFD) systems. While aerial communication using millimeter wave communication is an emerging field that offers high throughput, low latency, and low co-UAV interference \cite{6,7,8}, its typical coverage of only several hundred meters limits its applicability in long-distance use-cases requiring spans up to kilometers.

In-band full-duplex (IBFD) communication systems present a promising avenue for enhancing spectrum efficiency and overall system performance, due to the reuse of the same radio resource for both uplink and downlink transmissions. The key enabler of IBFD is self-interference cancellation (SIC), a topic that has been the focus of extensive research \cite{9}. Analog SICs, such as non-linear cancellation in the RF front-end \cite{13}, and digital SICs, such as reference receiver-assisted cancellations \cite{14,15}, have been investigated. Various strategies have also been explored to suppress self-interference in propagation, including physical or directional separation of receive (Rx) and transmit (Tx) antennas \cite{10}, polarization \cite{11}, and phase separation \cite{12}. It should be noted that these propagation-domain isolation techniques target a single transceiver, whose Tx--Rx interference geometry is fixed by the hardware layout and can thus be engineered offline at the design stage; moreover, in practice they serve as only the first stage of a multi-stage SIC chain and still need to be complemented by analog and digital cancellations. Because of the benefits of IBFD systems, they have also been incorporated into UAV-based applications, such as full-duplex UAV relays \cite{16,17,18} and UAV-mounted 5G base stations \cite{19}. However, the dedicated hardware, complicated algorithms, and additional energy required for SIC pose significant challenges. The challenges are particularly severe for UAV communication systems, which are subject to the UAV's stringent weight, size, and energy constraints. Currently, most of these systems are either direct adaptations or simplified versions of IBFD communication systems for terrestrial applications. There is a gap in the design of UAV-specific systems that take into account the characteristics of UAV platforms and aerial propagation environments. 

Experiments have also been carried out on aerial communication systems. A mmWave-based UAV communication system was developed for transmitting raw 4K video from a UAV to GS \cite{c17}. A study presented a prototype UAV controlled via an LTE link, examining the feasibility of using existing LTE infrastructure for UAV communication \cite{c15}. A testbed was built for UAV-to-car communication to evaluate performance with different UAV positions and antenna settings \cite{c14}. Field measurements of UAVs using LTE were conducted, with performance analyzed through simulations for large-scale UAV deployments \cite{c16}. A comprehensive overview of various prototypes and experimental efforts in UAV communications is detailed in \cite{c18}.

In our earlier works \cite{c12,c13,exp}, we introduced the concept of a multi-UAV IBFD (MU-IBFD) communication system, in which full-duplex operation is achieved at the system level rather than within a single transceiver, and reported preliminary prototypes and field trials. Those works demonstrated that, by equipping both the GS and UAVs with high-gain, beam-steerable directional antennas and by controlling the UAV geometry, self-interference can be converted into manageable co-channel interference (CCI) between UAVs, and enables full-duplex operation without heavyweight SIC on UAVs. 
Exploiting antenna directivity and spatial separation to mitigate interference has appeared in the literature \cite{9,10,11,12}, but it has been employed mainly as an auxiliary passive-isolation stage inside a single static transceiver, where the achievable isolation is a fixed hardware parameter and the residual self-interference is still left to analog and digital cancelers. This paper is fundamentally different: the geometry is elevated from an auxiliary isolation technique to the primary architectural mechanism, with no on-board canceler behind it. Realizing this in a multi-UAV air-to-ground network raises new challenges. First, the interference geometry is no longer fixed by the hardware layout but varies continuously with the flight of the UAVs, so the achievable isolation becomes a function of the time-varying relative UAV positions and must be explicitly modeled and guaranteed over the whole operating airspace. Second, the inter-UAV channel is almost persistently LOS at typical flight altitudes, so the interference can only be managed through the antenna patterns and the geometry itself. Third, the stringent weight, size, and energy constraints of UAV platforms preclude the multi-stage SIC chains of conventional IBFD transceivers, calling for an architecture-level rather than hardware-level solution.

Building on our earlier conceptual studies, this paper addresses the above challenges and advances the MU-IBFD concept to a complete, analytically characterized, and experimentally verified system design. The main contributions of this paper are summarized as follows:
\begin{itemize}
\item \textbf{Geometry-aware MU-IBFD architecture:} A complete architecture in which paired uplink/downlink channels are cross-reused within each UAV pair, transforming the intractable self-interference of conventional IBFD into inter-UAV CCI controllable by the UAV geometry, without any on-board canceler.
\item \textbf{Interference model and geometric design rule:} A geometry-aware CCI model and a reliable operating region (ROR) condition, which translate a target SINR into an explicit constraint on the relative 3D positions of the UAVs, directly usable by the flight controller.
\item \textbf{Prototype and field verification:} A two-UAV/one-GS prototype and kilometer-scale field trials, in which the measured CCI distribution agrees well with the predicted non-ROR region and shows that comparable performance would require ultra-high-performance on-board SI cancellation in conventional IBFD.
\item \textbf{Proof-of-concept demonstration:} A simultaneous dual 4K/60p video transmission over 5\,km, verifying the end-to-end practicality of the proposed system.
\end{itemize}

The rest of this paper is organized as follows. Section~II develops the complete MU-IBFD architecture and the associated interference model. Section~III describes the prototype implementation and experimental setup. Section~IV presents and analyzes the measurement results, including ROR verification, spectral-efficiency evaluation, and a proof-of-concept 4K/60p video demonstration. Section~V discusses practical considerations beyond the experimental conditions. Finally, Section~VI concludes the paper.

\section{Geometry-aware Multi-UAV Full-Duplex Communication: System Architecture}

To assist the readers in understanding the experiment, the architecture of MU-IBFD is recalled in this section. It is also noted that the target application of this research is an air-to-ground (multiple UAVs to GS) communication system, and the system topology of interest is star topology rather than peer-to-peer or line topology.

\subsection{Conventional IBFD Architecture}

\begin{figure}
    \centering
    \subfloat[Conventional IBFD architecture]{\includegraphics[width=.45\textwidth]{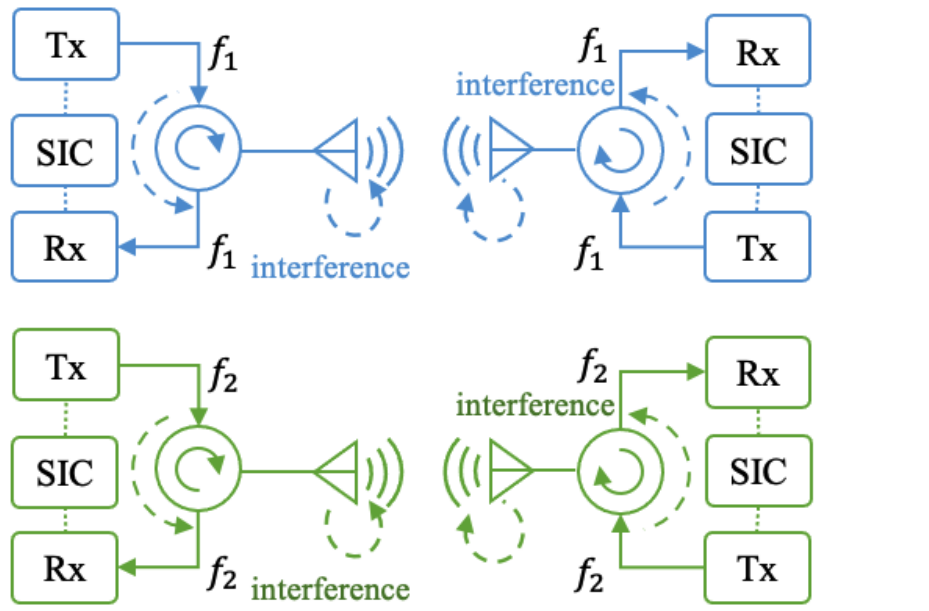}
    \label{fig:cibfd}}
    \hfil
    \centering
    \subfloat[Two-UAV IBFD]{\includegraphics[width=.45\textwidth]{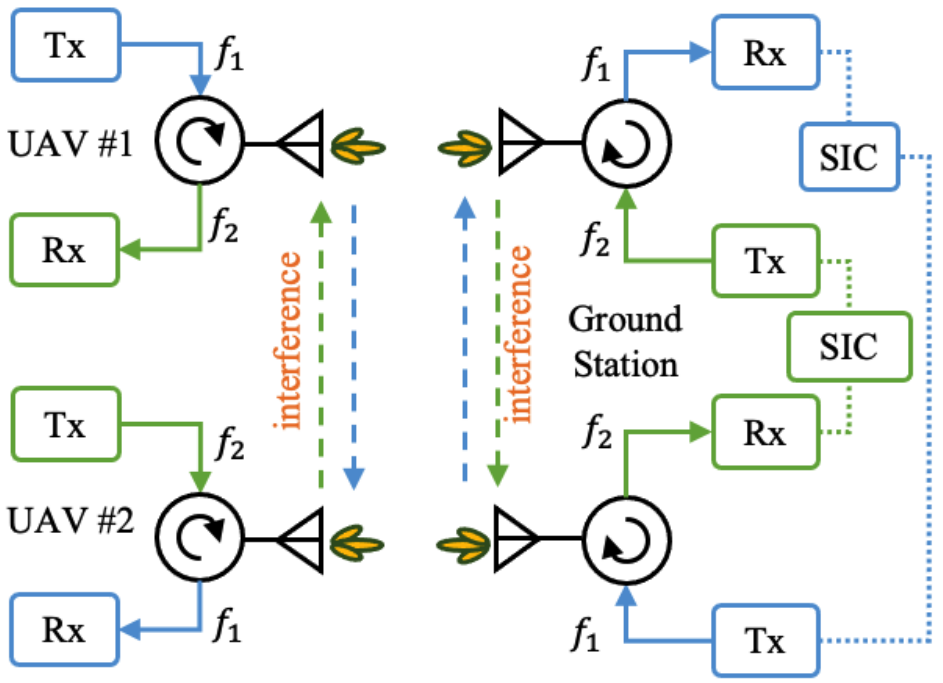}
    \label{fig:two-uav-ibfd}}

    \caption{Conventional single-node IBFD transceiver and proposed two-UAV IBFD architecture}
    \label{fig:ibfd}
\end{figure}

In a typical architecture of conventional IBFD transceivers shown in Fig.~\ref{fig:cibfd}, because wireless links in bi-directions share fully overlapping time and spectrum resources, the radio resource efficiency of IBFD systems is theoretically up to more than two times that of OBFD systems, in which radio resources cannot be re-used and extra guard intervals or bands must be added to avoid interference.
In IBFD, the large and inevitable self-interference from Tx chain to Rx chain in the same transceiver becomes the main issue in fostering IBFD. Self-interference is typically caused by a variety of factors, such as leakages in circulators, mismatching of antennas, and loopback signals in near fields.
Usually, self-interference does not only have significantly higher power than desired Rx signals (which could be up to 100 dB larger), but also contains a great deal of linear and non-linear distortion. Therefore, SIC must be performed to help extract the extremely weak Rx signals of interest from the distorted composite signals to enable error-free communications. A series of functional modules are necessary to be introduced between Tx and Rx chains for SIC in different domains, such as passive Tx-Rx isolation in the propagation domain, analog canceler in the analog domain, and digital canceler in the baseband domain. To implement a high-performance SIC, hardware complexity and energy costs would be greatly increased compared to OBFD or semi-duplex systems. Therefore, the conventional IBFD architecture is not suitable for UAVs, which are highly weight, size, and energy constrained.

\subsection{Two-UAV IBFD Architecture}
To achieve a UAV communication system featuring high performance and resource-efficiency while maintaining low hardware complexity, two conventional IBFD links can be extended to a two-UAV IBFD architecture, as shown in Fig.~\ref{fig:two-uav-ibfd}. Each UAV utilizes two distinct and separate channels for the link from a UAV to GS (hereafter referred to as uplink) and the link from GS to a UAV (hereafter referred to as downlink), thereby circumventing the need for complex hardware for SIC inherent in conventional IBFD for UAVs. Meanwhile, to enhance spectrum efficiency, each channel is reallocated to the uplink of one UAV and the downlink of the other. For instance, as depicted in Fig.~\ref{fig:two-uav-ibfd}, Channel\#$1$ with carrier frequency $f_1$ (blue) is reused by the uplink of UAV\#$1$ and the downlink of UAV\#$2$. In the same way, Channel\#$2$ with carrier frequency $f_2$ (green) is reused by the downlink of UAV\#$1$ and the uplink of UAV\#$2$. In this two-UAV IBFD architecture, each channel is concurrently reused by an uplink and a downlink, so in this two-UAV system, IBFD communication is achieved.
The full-duplex capability of this architecture lies at the system level. From the viewpoint of a single UAV, the uplink and the downlink work on two separate carriers as in FDD. At the channel level, however, every channel carries an uplink and a downlink simultaneously, so the whole band is occupied bi-directionally, as in an ideal IBFD system. The architecture thus attains the spectrum utilization of IBFD with only the on-board hardware complexity of FDD, and the equivalent cancellation metric $C^{\rm eq}$ introduced below quantifies this equivalence.

In this architecture, because self-interference between Tx and Rx in the same UAV is transformed into co-channel interference between a pair of two UAVs, which reuse the same two channels, dedicated SICs in UAVs for downlinks are not needed. On one UAV, uplink and downlink use separate channels, allowing leakage between Tx and Rx chains to be effectively suppressed by simple analog band-pass filters. Such an architecture is much simpler than the SIC in conventional IBFD and is more suitable for UAV-based systems. Because typically there are no special physical or energy constraints for GS as in UAVs, the architecture, such as SICs in GS, for uplinks can adopt the same hardware and algorithms as in conventional IBFD.

\begin{figure}
    \centering
    \includegraphics[width=.48\textwidth]{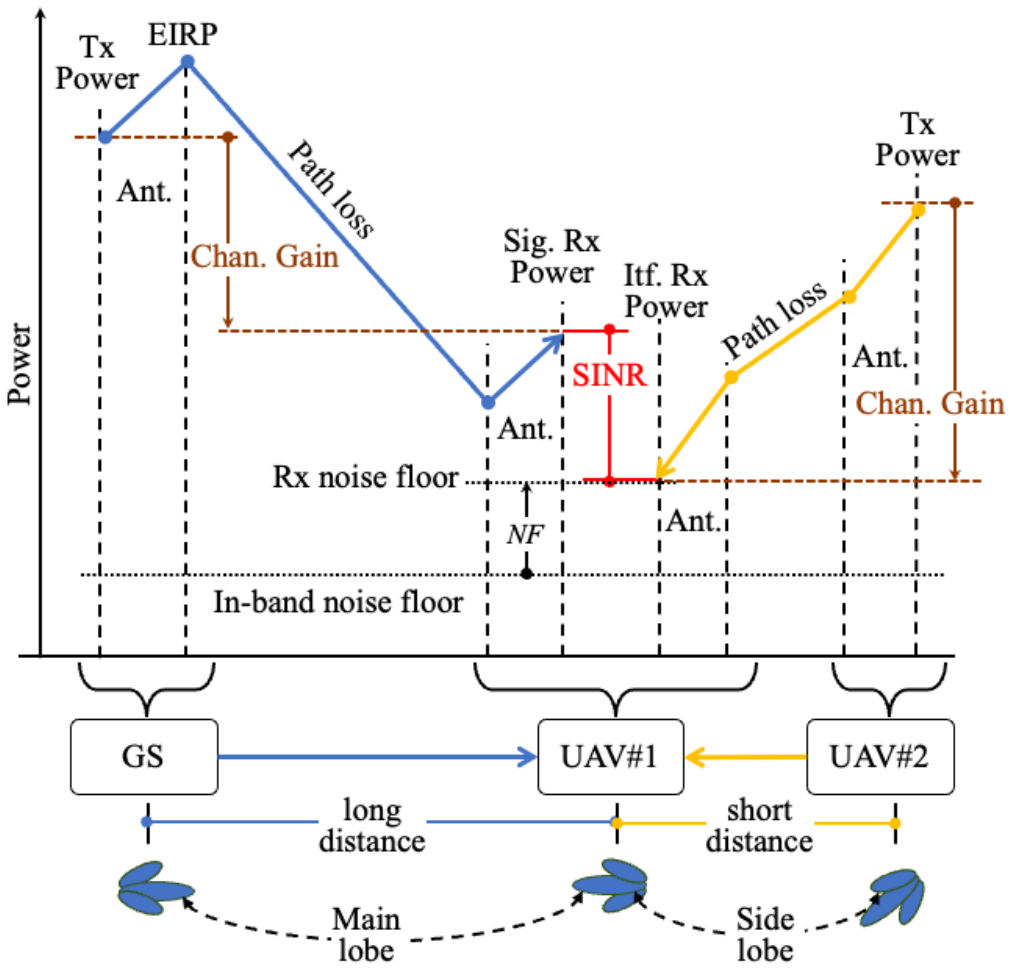}
    \caption{Power levels of signal of interest (blue) and interference (yellow) in the downlink of UAV\#1}
    \label{fig:powerlevel}
\end{figure}

Taking the downlink of UAV\#1 as an example, denote by $x^{\rm DL}_1(t)$ the downlink signal transmitted from the GS on this channel and by $x^{\rm UL}_2(t)$ the uplink signal from UAV\#2 on the same channel. The received baseband signal at the downlink Rx of UAV\#1 can be expressed as:
\begin{equation}
y_1(t) 
= h_{G\to 1}\,x^{\rm DL}_1(t)
+ h_{2\to 1}\,x^{\rm UL}_2(t)
+ w_1(t)
\label{eq:twoUAV_signal_model}
\end{equation}
where $h_{G\to 1}$ and $h_{2\to 1}$ are the complex channel gains of the GS-to-UAV\#1 downlink and the UAV\#2-to-UAV\#1 interference link, respectively, and $w_1(t)$ denotes the receiver noise. 
The corresponding average powers of them are:
\begin{equation}
\begin{aligned}
P_1^{\rm S}
  &= \mathbb{E}\{|h_{G\to 1} x^{\rm DL}_1(t)|^2\}
   = P_{\rm DL}\,|h_{G\to 1}|^2 \\
P_1^{\rm CCI}
  &= \mathbb{E}\{|h_{2\to 1} x^{\rm UL}_2(t)|^2\}
   = P_{\rm UL}\,|h_{2\to 1}|^2 \\
N_{\rm Tot}
  &= \mathbb{E}\{|w_1(t)|^2\}
\end{aligned}
\label{eq:power_gain_form}
\end{equation}
which correspond to the blue, yellow and noise power levels in Fig.~\ref{fig:powerlevel}. 
Here $P_{\rm DL} \triangleq \mathbb{E}\{|x^{\rm DL}_1(t)|^2\}$ and $P_{\rm UL} \triangleq \mathbb{E}\{|x^{\rm UL}_2(t)|^2\}$ denote the transmit powers on the downlink and the uplink of UAV\#2, respectively.
Under a quasi-static channel within each transmission block and assuming that the data symbols are independent of the channel, the equalities in \eqref{eq:power_gain_form} express the received powers as the product of the transmit powers and the effective channel power gains $|h_{G\to 1}|^2$ and $|h_{2\to 1}|^2$, which include both path-loss and antenna directivity. 

The key factor affecting performance is thus the CCI term in \eqref{eq:twoUAV_signal_model}, i.e., the interference from the uplink of one UAV to the downlink of the other. 
The resulting downlink signal-to-interference ratio (SIR) at UAV\#1 is:
\begin{equation}
{\rm SIR}_1
= \frac{P_1^{\rm S}}{P_1^{\rm CCI}}
= \frac{P_{\rm DL}}{P_{\rm UL}}
  \frac{|h_{G\to 1}|^2}{|h_{2\to 1}|^2}.
\label{eq:sir_basic}
\end{equation}
By equipping both the GS and UAVs with high-gain, beam-steerable directional antennas and steering their main beams towards the GS, the effective desired-link gain $|h_{G\to 1}|^2$ is kept close to its maximum, while the interference link UAV\#2 $\to$ UAV\#1 is confined to side- or back-lobes so that $|h_{2\to 1}|^2$ becomes much smaller. Consequently, the ratio in \eqref{eq:sir_basic} is significantly increased, which corresponds to the large separation between the desired-signal (blue) and interference (yellow) power levels in Fig.~\ref{fig:powerlevel}.
Moreover, the 3D-space maneuverability of UAVs is exploited to further reduce co-channel interference. Active maneuvering of UAVs can be performed according to the antenna directivities and positional relations of two UAVs.

To relate the proposed architecture to conventional single-node IBFD transceivers, we introduce the equivalent SI cancellation requirement. Let $P_{\mathrm{SI,0}}$ denote the SI power at the receiver input of a
conventional IBFD transceiver before SIC, and let $P_{\mathrm{SI,res}}$ be the residual SI power after SIC, so the SIC provides a suppression factor $C=P_{\mathrm{SI,0}}/ P_{\mathrm{SI,res}}$.
The proposed two-UAV IBFD architecture is said to be
equivalent to a conventional IBFD transceiver with cancellation
factor $C_{\mathrm{eq}}$ when the residual self-interference power equals the resulting CCI at UAV\#1, i.e., 
$P_{\mathrm{SI,res}} = P_{\mathrm{CCI}}$, so it derives:
\begin{equation}
  C^{\mathrm{eq}}
  = \frac{P_{\mathrm{SI,0}}}{P_{\mathrm{SI,res}}}
  = \frac{P_{\mathrm{SI,0}}}{P_{\mathrm{CCI}}}
  \label{eq:Ceq}
\end{equation}
In other words, the geometry- and antenna-induced reduction
of CCI in the two-UAV IBFD architecture plays the same role
as a SIC that suppresses the SI by $C^{\mathrm{eq}}$ in a conventional IBFD transceiver.

\subsection{MU-IBFD Architecture}
\begin{figure*}
    \centering
    \includegraphics[width=.95\textwidth]{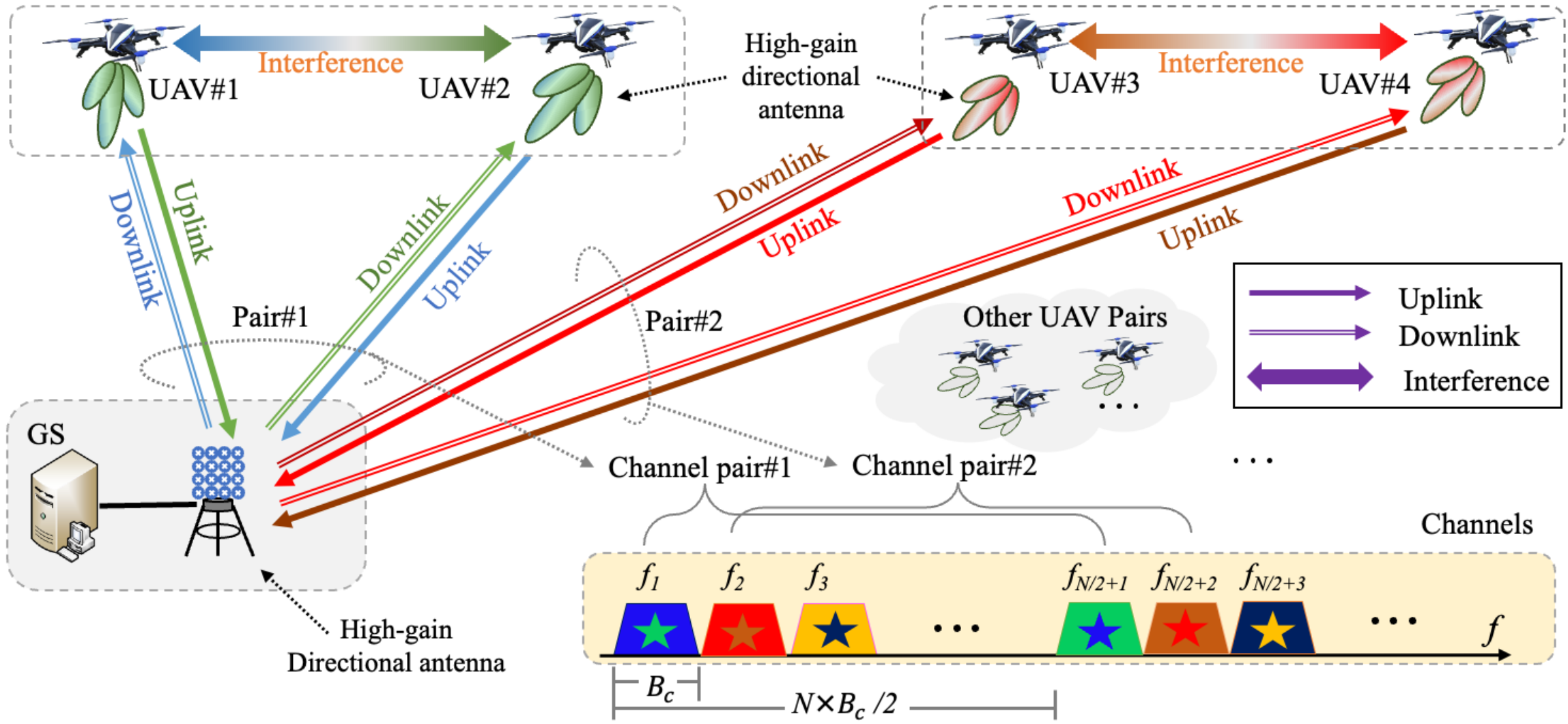}
    \caption{Architecture of the proposed geometry-aware MU-IBFD system and channel allocation with paired uplink/downlink channels}
    \label{fig:mu-ibfd}
\end{figure*}

As depicted in~Fig.~\ref{fig:mu-ibfd}, to maximize bandwidth utilization, we further extend this two-UAV IBFD architecture to a multi-UAV IBFD (MU-IBFD) system for multiple UAVs to communicate bi-directionally and simultaneously with the GS via high-performance full-duplex links. Each UAV pair reuses two separate channels for uplinks and downlinks, and different UAV pairs are assigned non-overlapping channel pairs to eliminate interference between different pairs.

The following channel allocation scheme is employed to achieve sufficient spectrum separation between the uplink and downlink of each UAV, as shown in Fig.~\ref{fig:mu-ibfd}. The full band is divided into $N$ channels, i.e., Channel\#$1, 2, \ldots, N$, with bandwidth $B_c$ and carrier frequencies $f_1,f_2, \ldots,f_N$. 
For convenience, let the total number of channels $N$ be even and define
$\mathcal{K} \triangleq \{1,2,\dots,N/2\}$ as the index set of channel pairs.
For each $k \in \mathcal{K}$, the two channels are grouped into one  pair:
\begin{equation*}
\bigl(\text{Channel}\#k,\; \text{Channel}\#(N/2+k)\bigr)
\end{equation*}
Hence, up to $|\mathcal{K}| = N/2$ UAV pairs can be supported simultaneously in the MU-IBFD system.

An example is also given in Fig.~\ref{fig:mu-ibfd}. Channel\#$1$ (marked as blue) and Channel\#$N/2+1$ (marked as green) are a pair of two channels, of which Channel\#$1$ is assigned to UAV\#$1$ downlink and UAV\#$2$ uplink; Channel\#$N/2+1$ is assigned to UAV\#$1$ uplink and UAV\#$2$ downlink. Channel\#$2$ (marked as red) and Channel\#$N/2+2$ (marked as brown) are assigned to UAV\#$3$ and UAV\#$4$ following the same principle. 
More generally, let the $m$-th UAV pair consist of UAV\#$(2m-1)$ and UAV\#$(2m)$,
and let it use the channel pair $(k_m, N/2+k_m)$.
Let $c^{\rm UL}_u$ and $c^{\rm DL}_u$ denotes the channel indices used by UAV\#$u$ for uplink and downlink, respectively. Then, for the $m$-th UAV pair, we have:
\begin{equation}
\begin{aligned}
&c_{2m-1}^{\rm UL} = N/2 + k_m, &\quad &c_{2m}^{\rm DL}   = N/2 + k_m\\
&c_{2m}^{\rm UL}   = k_m, &\quad &c_{2m-1}^{\rm DL} = k_m
\end{aligned}
\label{eq:MU_channel_mapping}
\end{equation}

By using such channel division and assignment principles, the spectrum separation between uplink and downlink up to half of the full band, $B_c N/2$, can be achieved on each UAV. Hence, adjacent-channel interference (ACI) between uplink and downlink on one UAV can be eliminated by analog filters and is thus considered negligible in later sections of this paper.
Although the system is extended to a multi-UAV architecture, the dominant factor affecting performance is still co-channel interference within a pair of two UAVs in downlinks. 
Since different UAV pairs are assigned disjoint channel pairs $\bigl(\text{Channel}\#k_m,\text{Channel}\#(N/2+k_m)\bigr)$, there is no co-channel interference between different pairs, so CCI only appears within each UAV pair, as in the two-UAV case.
As explained above, high-gain directional beam-steerable antennas and the active maneuverability of UAVs can be utilized to eliminate co-channel interference and enable MU-IBFD communications.

The channel-pairing principle also determines how the system scales. Since no CCI exists between different UAV pairs, the $N$-UAV system decomposes exactly into $N/2$ mutually independent UAV pairs: the CCI within each UAV pair follows exactly the same mechanism as in the two-UAV case regardless of the total number of UAVs, and the system capacity scales linearly with the number of UAV pairs until the total bandwidth $B_c N$ is exhausted. The only residual coupling between different UAV pairs is the ACI caused by adjacent-channel leakage. Compared with the CCI at the same relative geometry, the ACI power is further reduced by the total adjacent-channel suppression $A_{\rm ACL}$, i.e., $P^{\rm ACI}\approx P^{\rm CCI}/A_{\rm ACL}$, where $A_{\rm ACL}$ combines the transmitter adjacent-channel leakage ratio (ACLR) and the receiver selectivity, and typically amounts to several tens of dB, so the ACI is negligible even at positions where the CCI itself is critical. Consequently, the two-UAV configuration analyzed above is not a simplification but the elementary unit of the proposed architecture, and its analysis and experimental verification in the following sections carry over to any number of UAV pairs without loss of generality. The scalability bottleneck is the amount of available spectrum, a resource-allocation problem common to all multi-UAV systems \cite{refC}, rather than interference management. It is worth noting that if more than two UAVs were to share one channel pair, the CCI would couple multiple links simultaneously, and multi-UAV simulation and scheduling would then become necessary. Such many-to-one channel reuse could further improve the spectrum efficiency, but it is a different architecture from the pairing principle adopted here and is left for future work.

\subsection{Reliable Operating Region}

\subsubsection{Geometry Preliminaries}

Because each UAV mechanically steers its antenna towards the GS, the geometry of the interference link is completely determined by the UAV positions. We introduce a global Cartesian coordinate system in which the GS is at the origin, and the positions of the two UAVs are $\boldsymbol r_1 = (x_1,y_1,z_1)$ and $\boldsymbol r_2 = (x_2,y_2,z_2)$. The vector from the GS to UAV $i$ is $\boldsymbol r_i$ with distance $D_i = \|\boldsymbol r_i\|$. The vector from UAV\#2 to UAV\#1 is $\boldsymbol v = \boldsymbol r_1 - \boldsymbol r_2$ with $d = \|\boldsymbol v\|$, and the main-beam direction of the antenna on UAV $i$ (pointing towards the GS) is represented by the unit vector $\boldsymbol b_i = -\boldsymbol r_i / D_i$. Let $\hat{\boldsymbol v} = \boldsymbol v/d$ denote the unit vector from UAV\#2 to UAV\#1.
In the general case of arbitrary UAV positions, the azimuth and elevation offsets of the interference link are obtained by expressing $\hat{\boldsymbol v}$ (at the Tx side, UAV\#2) and $-\hat{\boldsymbol v}$ (at the Rx side, UAV\#1) in the local antenna coordinate system whose boresight axis is aligned with the corresponding $\boldsymbol b_i$. Since each $\boldsymbol b_i$ is determined by $\boldsymbol r_i$ through the GS-pointing steering rule, the geometry model itself imposes no restriction on the UAV positions.

In the typical long-distance operating scenarios targeted in this paper, the inter-UAV distance is much smaller than the GS--UAV distance, i.e., $d \ll D_i$, which allows the angle calculation to be simplified as follows. The GS is located far away along the $x$-axis. Both UAVs are at approximately the same distance $D$ from the GS, and their antenna boresights are nearly parallel and directed towards the GS. Under these conditions, $\boldsymbol b_1 \approx \boldsymbol b_2 \approx \boldsymbol b$.
If the UAVs are located in the region $x>0$, the boresight direction can be approximated by $\boldsymbol b \approx (-1,0,0)$.
Let the relative displacement between the UAVs be
$\Delta x=x_1-x_2$,
$\Delta y=y_1-y_2$,
$\Delta h=z_1-z_2$,
and define
$d = \sqrt{\Delta x^2+\Delta y^2+\Delta h^2}$,
$\hat{\boldsymbol v}=(\Delta x,\Delta y,\Delta h)/{d}$.
The unit vector $\hat{\boldsymbol v}$ represents the direction from UAV\#2 to UAV\#1. Expressed in the local antenna coordinates whose $x'$-axis aligns with $\boldsymbol b$, the azimuth and elevation offsets of the interference link at UAV\#2 (Tx) and UAV\#1 (Rx) can be calculated as
$\phi_{\rm Tx} = \operatorname{atan2}\!\left(\Delta y, -\Delta x\right)$,
$\phi_{\rm Rx} = \phi_{\rm Tx} + \pi$,
$\psi_{\rm Tx} = \operatorname{atan2}\!\left(\Delta h,\sqrt{\Delta x^2+\Delta y^2}\right)$,
$\psi_{\rm Rx} = -\psi_{\rm Tx}$.
The accuracy of this approximation is governed by the dimensionless ratio $d/D$: the angular deviation between the two boresights is at most approximately $d/D$ (in radians), which is negligible as long as the inter-UAV distance is much smaller than the GS--UAV distance, i.e., whenever $d/D$ is small compared with the antenna HPBW. This is the typical operating regime of the target applications, in which paired UAVs fly in each other's vicinity while communicating with a GS several kilometers away (e.g., in the field trials described in Section~III, $d\approx100$\,m and $D=2.5$\,km yield a deviation of about $2.3^{\circ}$, far below the $27.5^{\circ}$ HPBW of the UAV antennas). It should be emphasized that the ROR formulation developed below is defined for arbitrary UAV positions, and the approximation is introduced only to obtain compact closed-form angle expressions for ease of use in practice.

\subsubsection{Reliable Operating Region (ROR)}
Within a pair of UAVs, CCI from one UAV's uplink Tx to the other's downlink Rx can be written as (taking interference at UAV\#1 as an example)
\begin{equation}
\label{eq:cci}
P_{1}^{\text{CCI}}
=
P_{2}^{\text{Tx}}
P_{{ 2} \rightarrow { 1}}^{\text{C}}
\end{equation}
where UAV\#1 and UAV\#2 are two UAVs reusing the same channels, $P_{ 1}^{\text{CCI}}$ is the interference received at the downlink Rx of UAV\#1, $P_{ 2}^{\text{Tx}}$ is the uplink Tx power of UAV\#2, and $P_{{ 2} \rightarrow { 1}}^{\text{C}}$ is the channel power gain from UAV\#2 to UAV\#1, including both antenna gains and wireless channel effects.

For the downlink of UAV\#1, given the relative positions $(\boldsymbol r_{1}, \boldsymbol r_{2})$ of the two UAVs, the received signal-to-interference-and-noise ratio (SINR) as:
\begin{equation}
\label{eq:sinr1}
\mathrm{SINR}_1
=
\frac{P_{1}^{\rm S}}
     {P_{1}^{\rm CCI} + N_{\rm Tot}}
\end{equation}
where $P_{1}^{\rm S}$ is the received power of the desired downlink signal from the GS, $P_{1}^{\rm CCI}$ is the received co-channel interference power from the uplink of UAV\#2 as in \eqref{eq:cci}, and $N_{\rm Tot}$ denotes the total noise power at the downlink receiver, including thermal noise and receiver noise figure. Let $\gamma_{\rm min}$ be the target SINR required by a given modulation and coding scheme.

A pair of UAV positions $(\boldsymbol r_{1}, \boldsymbol r_{2})$ is said to be in the ROR if $\mathrm{SINR}_1 \ge \gamma_{\rm min}$.
Equivalently, this condition can be written as an upper bound on the allowable interference power:
\begin{equation}
\label{eq:Ith}
P_{1}^{\rm CCI}
\le
\underbrace{\frac{P_{1}^{\rm S}}{\gamma_{\rm min}} - N_{\rm Tot}}_{\displaystyle I_{\rm th}},
\end{equation}
where $I_{\rm th}={{P_{1}^{\rm S}}/{\gamma_{\rm min}} - N_{\rm Tot}}$ is the maximum tolerable interference power. 
In practical systems, the GS-to-UAV received power $P_{1}^{\rm S}(\boldsymbol{r}_{1})$ is routinely estimated from downlink reference signals, such as channel‑state information reference signals (CSI‑RS) and demodulation reference signals (DMRS), using standard reference signal received power (RSRP) measurements, and reported to the GS.
In the experiments, for a given measurement point, $P_{1}^{\rm S}$ can be accurately and directly measured by turning off UAV\#2. (Since it only appears through $I_{\rm th}$, an explicit analytical expression of the GS–UAV channel is not required in this section.)

Combining \eqref{eq:cci} and \eqref{eq:Ith} yields a general condition on the interference channel gain:
\begin{equation}
P_{{ 2} \rightarrow { 1}}^{\text{C}}
\bigl(\boldsymbol{r}_{2},\boldsymbol{r}_{1}\bigr)
\le
\frac{I_{\rm th}}{P_{ 2}^{\text{Tx}}}
\label{eq:general_safe}
\end{equation}
All UAV position pairs $(\boldsymbol r_{1}, \boldsymbol r_{2})$ that satisfy \eqref{eq:general_safe} belong to the ROR. 
Note that $P_{{ 2} \rightarrow { 1}}^{\text{C}} \bigl(\boldsymbol{r}_{2},\boldsymbol{r}_{1}\bigr)$ also depends on the steering rule (e.g., always pointing to the GS) and on the Tx/Rx antenna patterns, which are omitted in \eqref{eq:general_safe} because they are fixed for a given system.

\subsubsection{Condition for ROR}
In the typical target application scenario, the channel between the two UAVs is strongly LOS-dominant. The contributions of the remaining multi-path components (e.g., ground reflection) and modeling uncertainties can be captured by a shadowing margin.
Note that conditions in which the LOS component is partially or fully blocked only reduce the interference, so the LOS assumption between two UAVs provides a worst-case characterization of CCI. 
For compactness, we define the composite antenna gain of a UAV pair as:
\begin{equation}
\mathcal G(\boldsymbol r_1,\boldsymbol r_2)
=
G_{\rm Tx}\!\bigl(\phi_{\rm Tx}, \psi_{\rm Tx}\bigr)
G_{\rm Rx}\!\bigl(\phi_{\rm Rx}, \psi_{\rm Rx}\bigr)
\label{eq:G_pair}
\end{equation}
where $G_{\rm Tx}(\phi_{\rm Tx},\psi_{\rm Tx})$ and $G_{\rm Rx}(\phi_{\rm Rx},\psi_{\rm Rx})$ are the UAV uplink and downlink antenna power patterns evaluated at the azimuth/elevation offsets of the interference link.
Under the LOS assumptions above, the interference channel gain can be approximated by:
\begin{equation}
P^{\mathrm{C}}_{\mathrm{UAV2}\to\mathrm{UAV1}}
\approx
\eta_{\rm sh} \mathcal G(\boldsymbol r_1,\boldsymbol r_2) L_{\rm UU}(d)
\label{eq:Pc_LOS}
\end{equation}
where $L_{\rm UU}(d)$ denotes the nominal large-scale path-loss of the UAV-to-UAV link, and $\eta_{\rm sh} \ge 1$ is a shadowing margin (e.g., corresponding to $M_{\rm sh}$ dB) that accounts for residual multi-path components and other modeling mismatches (e.g., small-scale fluctuations and beam misalignment caused by UAV attitude jitter and mechanical pointing errors of the antenna rotator), so that \eqref{eq:Pc_LOS} safely upper-bounds the actual interference channel gain.

The large-scale path-loss is modeled by a power-law function:
\begin{equation}
L_{\rm UU}(d) = K d^{-\alpha}
\label{eq:pathloss_UU}
\end{equation}
where $\alpha \approx 2$ for LOS air-to-air links and $K = \bigl({\lambda}/{4\pi}\bigr)^2$ or a more general empirical constant. Substituting \eqref{eq:Pc_LOS} and \eqref{eq:pathloss_UU} into the general safe condition \eqref{eq:general_safe} yields:
\begin{equation}
\eta_{\rm sh}\,
\mathcal G(\boldsymbol r_1,\boldsymbol r_2)\,
K d^{-\alpha}
\le
\frac{I_{\rm th}}{P_{2}^{\text{Tx}}}
\end{equation}
Rearranging terms and defining
$ \beta = {I_{\rm th}}/({P_{2}^{\text{Tx}}\eta_{\rm sh}K})$,
we obtain the following ROR condition:
\begin{equation}
\mathcal G(\boldsymbol r_1,\boldsymbol r_2)
\le
\beta\, {\|\boldsymbol r_1 - \boldsymbol r_2 \|}^{\alpha}
\label{eq:safe_theta}
\end{equation}
which defines the reliable operating region in terms of UAV positions under the LOS air-to-air model. $\mathcal G(\boldsymbol r_1,\boldsymbol r_2)$ can be calculated numerically using real 3D antenna patterns.
The condition \eqref{eq:safe_theta} admits an intuitive geometric interpretation. Under the far-field geometry of Section~II-D-1, $\mathcal G$ depends only on the direction of the interference link; hence, for each direction, \eqref{eq:safe_theta} is violated only within the exclusion distance $d_{\rm excl} = \left(\mathcal G/\beta\right)^{1/\alpha}$. The non-ROR region around each UAV is therefore a bounded, antenna-pattern-shaped volume whose radial extent follows the composite gain: largest along the mutual boresight direction and shrinking rapidly off-axis. This turns the ROR into a simple keep-out-zone rule for the flight controller, analogous to the separation minima in aviation.
Equation~\eqref{eq:safe_theta} therefore provides geometry-aware design rules for operating the MU-IBFD system: given the system parameters and the desired SINR threshold, the flight controller can determine reliable relative positions of the UAVs directly in the full 3D space $(\boldsymbol r_1,\boldsymbol r_2)$.
In particular, since the condition is formulated in the full 3D space, altitude differences between the UAVs are naturally handled through the vertical component $\Delta h$ of the relative displacement, so the same design rule applies to UAVs operating at different heights, which is an essential aspect of interference management in dynamic 3D deployments \cite{refA}.

\subsubsection{Impact of UAV Attitude on Beam Pointing}
In practice, the beam pointing deviates from the ideal GS direction, since quadrotor UAVs continuously adjust their attitude against wind and turbulence and the mechanical rotator compensates such rotations with finite accuracy. The impact of the residual pointing error $\delta$ is twofold. On the desired GS link, the boresight gain loss grows only quadratically, as $\Delta G \approx 12\,(\delta/\theta_{\rm 3dB})^{2}$\,dB, and thus remains a second-order effect as long as $\delta$ is a small fraction of the beamwidth---a condition that can be ensured by choosing the beamwidth according to the achievable attitude-compensation accuracy. On the interference link, $\delta$ shifts the angles at which the patterns in \eqref{eq:G_pair} are evaluated, and the worst-case increase of $\mathcal G$ is exactly what the shadowing margin $\eta_{\rm sh}$ in \eqref{eq:Pc_LOS} absorbs, so misalignment only consumes margin, without changing the structure of \eqref{eq:safe_theta}. Moreover, since the CCI depends only on the relative UAV positions and antenna patterns, not on their speeds, high-mobility operation acts mainly through the pointing accuracy, while the spread-spectrum downlink further protects safety-related control signaling. For the prototype antenna ($\theta_{\rm 3dB}\approx27.5^{\circ}$, Section~III), $\delta=3^{\circ}$ costs only about $0.14$\,dB; the field measurements in Section~IV, which inherently contain wind-induced jitter (cf. Fig.~\ref{fig:updownrssi}), still agree well with the ROR predicted using a 3\,dB margin.
It should be noted that \eqref{eq:safe_theta} is derived from a deterministic air-to-air model with a fixed shadowing margin. In more general environments, the same framework can be extended to a probabilistic ROR by modeling the residual fluctuation of $P^{\mathrm{CCI}}_1$ (caused by small-scale propagation, attitude jitter, and pointing errors) as a log-normal random factor and defining the operating region via an outage-probability constraint on SINR, but such extensions are beyond the scope of this paper.

\section{Description of Experiment}

Experiments are designed and conducted to confirm if the proposed system can effectively eliminate downlink co-channel interference and enable systematic full-duplex communication. Because interference only occurs from uplinks to downlinks between two UAVs reusing the same channels, without loss of generality, a minimum prototype system consisting of one GS and two UAVs is practically developed, and other UAV pairs follow the exact same mechanism.

\subsection{Prototype UAVs and GSs}

We developed and implemented a multi-UAV system capable of long-distance (up to 5 km) 4K(3840×2160)/60p video transmission (uplink) and control (downlink) as one of the typical UAV applications. The prototype system consists of one GS and two UAVs. (Please note that since this is a prototype, the current size is large. A more compact one less than 0.8\,L has already been developed.)

\begin{figure}
    \centering
    \subfloat[Refitted UAVs equipped with experimental devices and payload box]{\includegraphics[width=.48\textwidth]{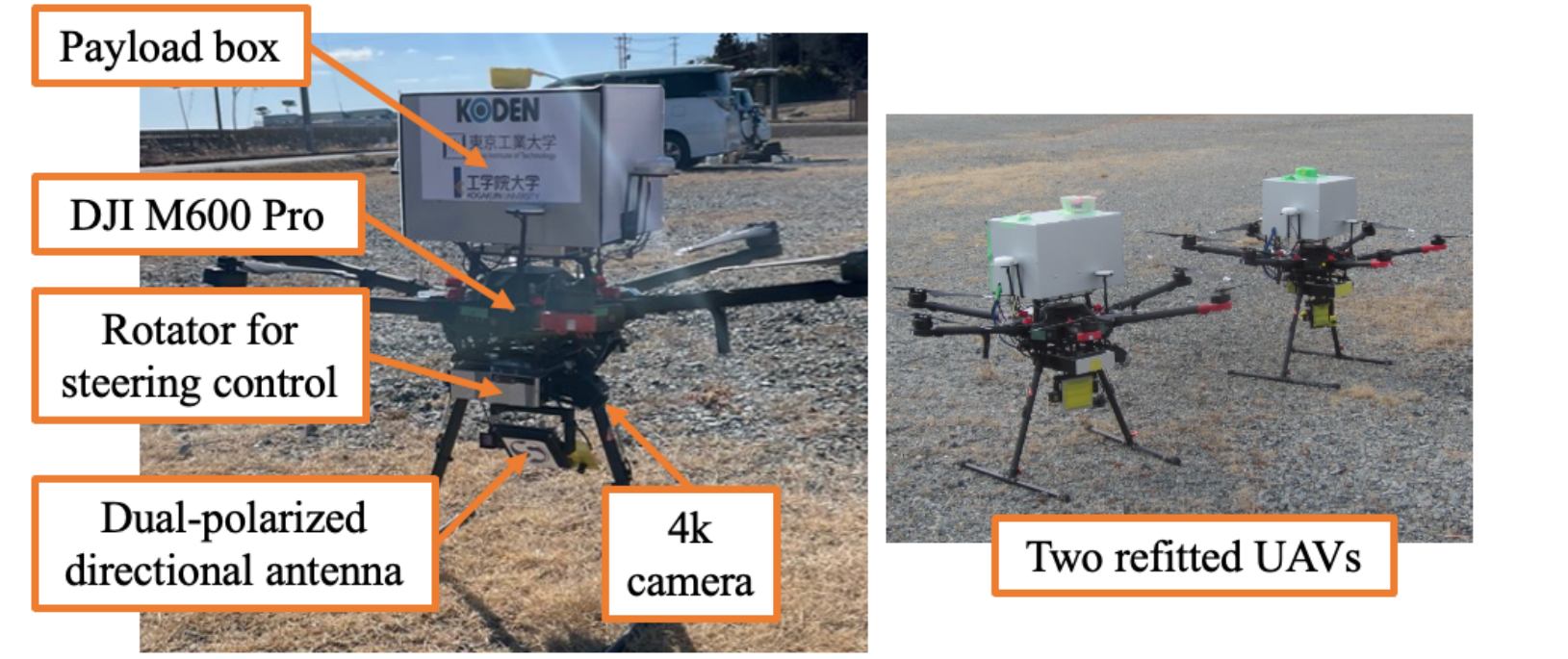}
    \label{fig:uava}}
    \hfill
    \subfloat[Transceiver (baseband processing and codec) in payload box]{\includegraphics[width=.48\textwidth]{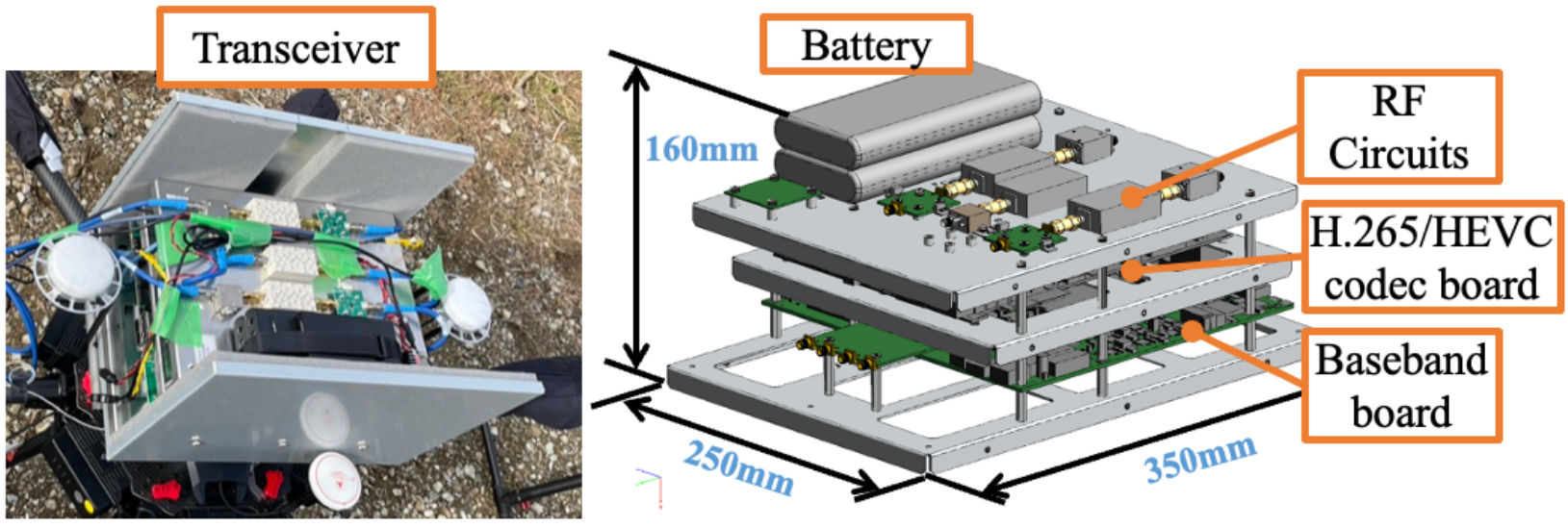}
    \label{fig:uavb}}
    \hfill
    \subfloat[UAV antenna and rotatory device for antenna steering control]{\includegraphics[width=.48\textwidth]{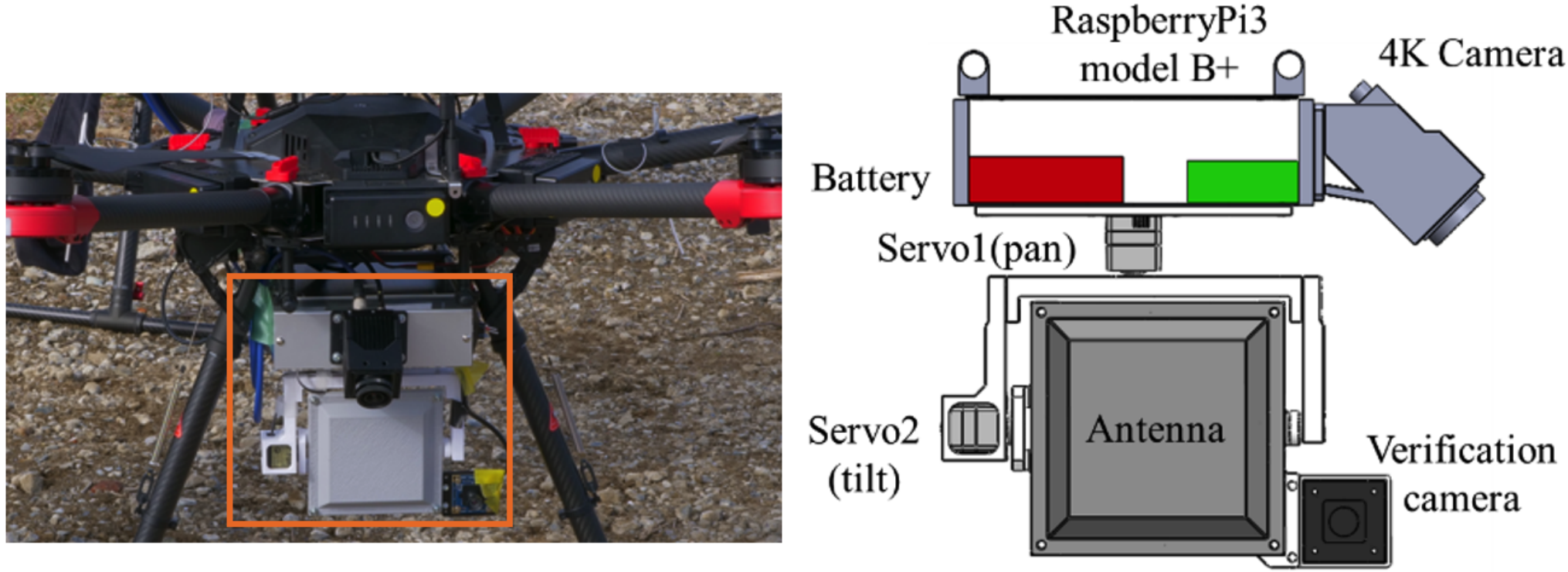}
    \label{fig:uavc}}
    \hfill
    \caption{Prototype UAV platform and on-board experimental equipment.}
    \label{fig:uav-prototype}
\end{figure}

\begin{figure}
    \centering
    \includegraphics[width=.45\textwidth]{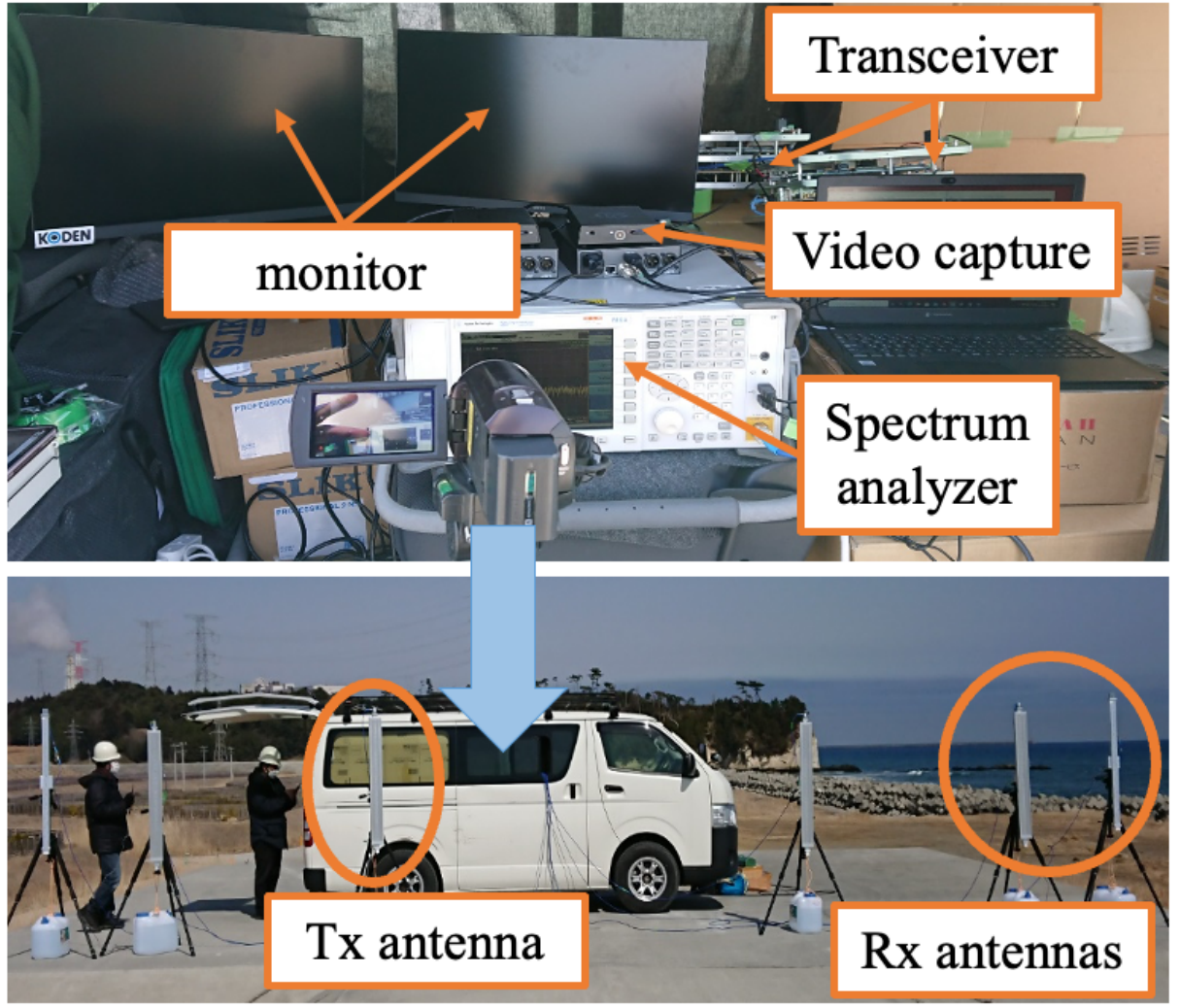}
    \caption{Prototype GS}
    \label{fig:gs}
\end{figure}

In this experiment, two UAVs were refitted from the DJI M600 Pro~\cite{dji} to carry all experiment equipment, including transceiver, 4K camera, dual-polarized directional antenna, antenna rotator, etc., as shown in Fig.~\ref{fig:uava}. The total payload weight is around 5 kg.
The prototype transceiver, including video codec, baseband processing (developed and implemented on FPGAs), and RF circuits, is installed in a payload box on the top of UAVs, as shown in Fig.~\ref{fig:uavb}. A directional antenna is installed on the downside of UAV. Mechanical beamforming is employed in this prototype, and the directional antenna is rotated mechanically by a 2-dimension (tilt and pan) rotator, which is controlled by a Raspberry Pi according to GS position (preset) and UAV positions (got from RTK-GPS in real-time), as shown in Fig.~\ref{fig:uavc}. A lightweight camera is attached to the antenna to confirm the antenna's directivity. The prototype GS employs conventional architecture, and all experiment devices, including transceivers, spectrum analyzer, monitors, video captures, etc., are installed in a vehicle, as depicted in Fig.~\ref{fig:gs}.

\begin{figure}
    \centering
    \includegraphics[width=.48\textwidth]{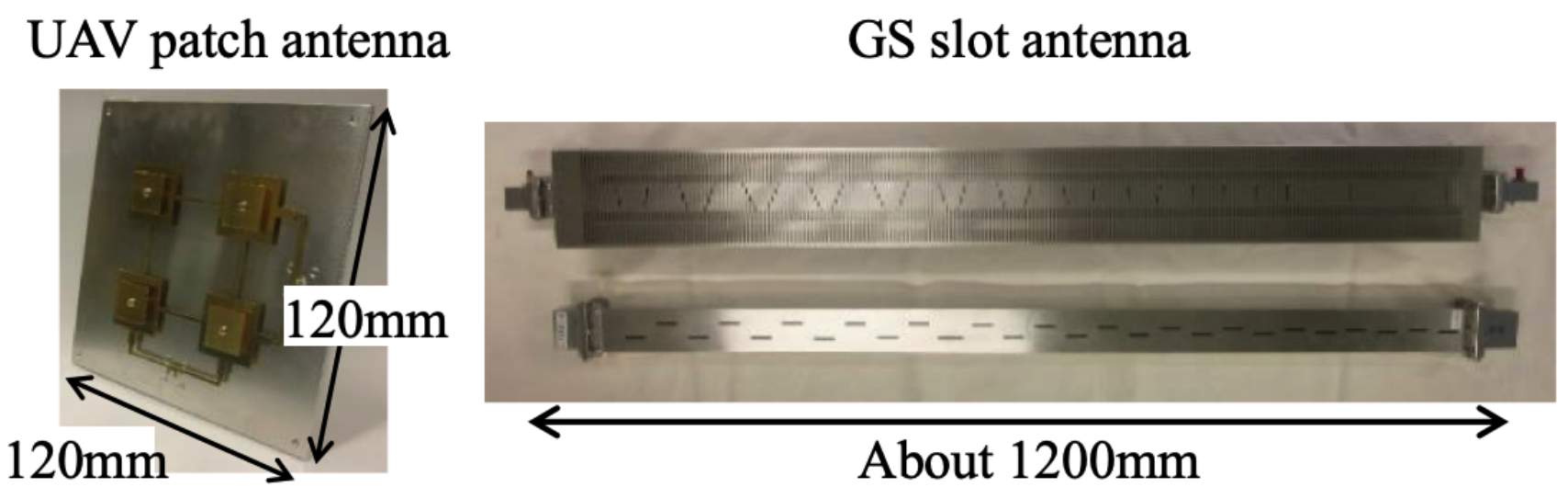}
    \caption{Prototype UAV patch antenna and GS slot antenna}
    \label{fig:antenna}
\end{figure}

High-gain directional antennas are practically employed in both UAVs and GS, as shown in Fig.~\ref{fig:antenna}. 
Dual-polarized patch antennas, with 15.3 dBi antenna gain, 27.5$^{\circ}$/29.2$^{\circ}$ and 27.4$^{\circ}$/28.4$^{\circ}$ half-power-beam-width (HPBW) (elevation/azimuth) in vertical and horizontal polarization, respectively, and 120 mm $\times$ 120 mm size, are fabricated for UAVs with joint consideration of high gain and small size/weight. In UAVs, antennas are shared by uplink Tx and downlink Rx. 
Since the size/weight of GS antennas are less restrictive than UAVs, slot antennas with bulky sizes, i.e., 1200 mm $\times$ 110 mm and 1150 mm $\times$ 70 mm, but high gain, i.e., 18.2 dBi and 19.8 dBi, a very narrow elevation beam (to reduce the ground reflection and increase the gain), and a wide azimuth beam (to cover UAVs typically with large horizontal movement), i.e., 3.0$^{\circ}$/52.0$^{\circ}$ and 3.2$^{\circ}$/75.0$^{\circ}$ HPBW (elevation/azimuth) in vertical and horizontal polarization, respectively, are used for GS. The parameters of antennas are summarized in Table~\ref{tb:antenna}.

\begin{table}
    \caption{Antenna parameters}
    \label{tb:antenna}
    \centering
    \begin{tabular}{lcc}
        
        \multicolumn{3}{c}{UAV patch antenna}  \\
        \hline \hline 
        & Ver. polarization & Hor. polarization \\
        \hline
        Antenna gain & 15.3 dBi & 15.3 dBi \\
        HPBW (el./az.) & 27.5$^{\circ}$ / 29.2$^{\circ}$ & 27.4$^{\circ}$ / 28.4$^{\circ}$ \\
        Size & \multicolumn{2}{c}{120mm$\times$120mm}  \\

        \hline 
        \\
        \multicolumn{3}{c}{GS slot antenna}  \\
        \hline \hline 
        & Ver. polarization & Hor. polarization \\
        \hline
        Antenna gain & 18.2 dBi & 19.8 dBi \\
        HPBW (el./az.) & 3.0$^{\circ}$ / 52.0$^{\circ}$ & 3.2$^{\circ}$ / 75.0$^{\circ}$ \\
        Size & 1200mm$\times$110mm & 1150mm$\times$70mm\\
        \hline 
  \end{tabular}
\end{table}

The prototype system works in 5.7 GHz band. The whole available bandwidth of 100 MHz is divided into 10 channels. Two channels with carrier frequencies of 5.675 GHz and 5.725 GHz (i.e., separation of 50 MHz), and bandwidth of 10 MHz (of which 1 MHz is used by guard band) are employed and shared by two uplinks and downlinks of a pair of two UAVs, as explained in the previous section.
The detailed design and parameters in the physical layer (such as modulation, channel coding, etc.) mainly reference the uplink in \cite{c19}. To enhance energy efficiency, single-carrier scheme is adopted. To ensure the stability of safety-related transmissions (such as flight control), spread spectrum is used in downlink. For high communication capacity and a compact size, the dual-polarization patch antenna and polarization-division multi-input-multi-output (MIMO) are used in UAVs. No interference cancelers are introduced in UAVs, except for band-pass filters to suppress ACI. GS employs conventional analog cancelers for IBFD.
Performance metrics such as channel power and bit error rate (BER) can be measured and logged in transceivers for performance evaluation. Detailed parameters are summarized in Table~\ref{tb:para}.

As explained above, different parameters (e.g., transmission power and modulation order) and techniques (e.g., multiple carrier and single carrier modulation) are deliberately adopted for the uplink and downlink design, due to the inherent asymmetry in this UAV application, where the uplink demands are significantly higher than downlink. It is noteworthy that consequently, the uplink and downlink will exhibit different performance metrics (e.g., received signal strength and SINR) in the experiment, but it is important to emphasize that such discrepancies do not affect the measurement of interference cancellation in the next section.

\begin{table}
    \caption{Parameters}
    \label{tb:para}
    \centering
    \begin{tabular}{lc}
        \multicolumn{2}{c}{Uplink (UAV to GS)} \\
        \hline \hline 
        Bandwidth & 10 MHz \\
        UAV Tx Power (ver./hor.)  & 17 dBm / 17 dBm \\
        Modulation  & 16QAM-SC \\
        Channel Coding & Turbo  \\
        Multiplexing  & Polarization MIMO \\
        \hline
        \\
        \multicolumn{2}{c}{Downlink (GS to UAV)} \\
        \hline \hline 
        Bandwidth & 10 MHz \\
        GS Tx Power & 8.75 dBm \\
        Modulation & QPSK-SC \\
        Channel Coding & Turbo \\
        Multiplexing & SISO \\
        \hline
  \end{tabular}
\end{table}

\subsection{Experiment Setup}

\begin{figure}
    \centering
    \includegraphics[width=.48\textwidth]{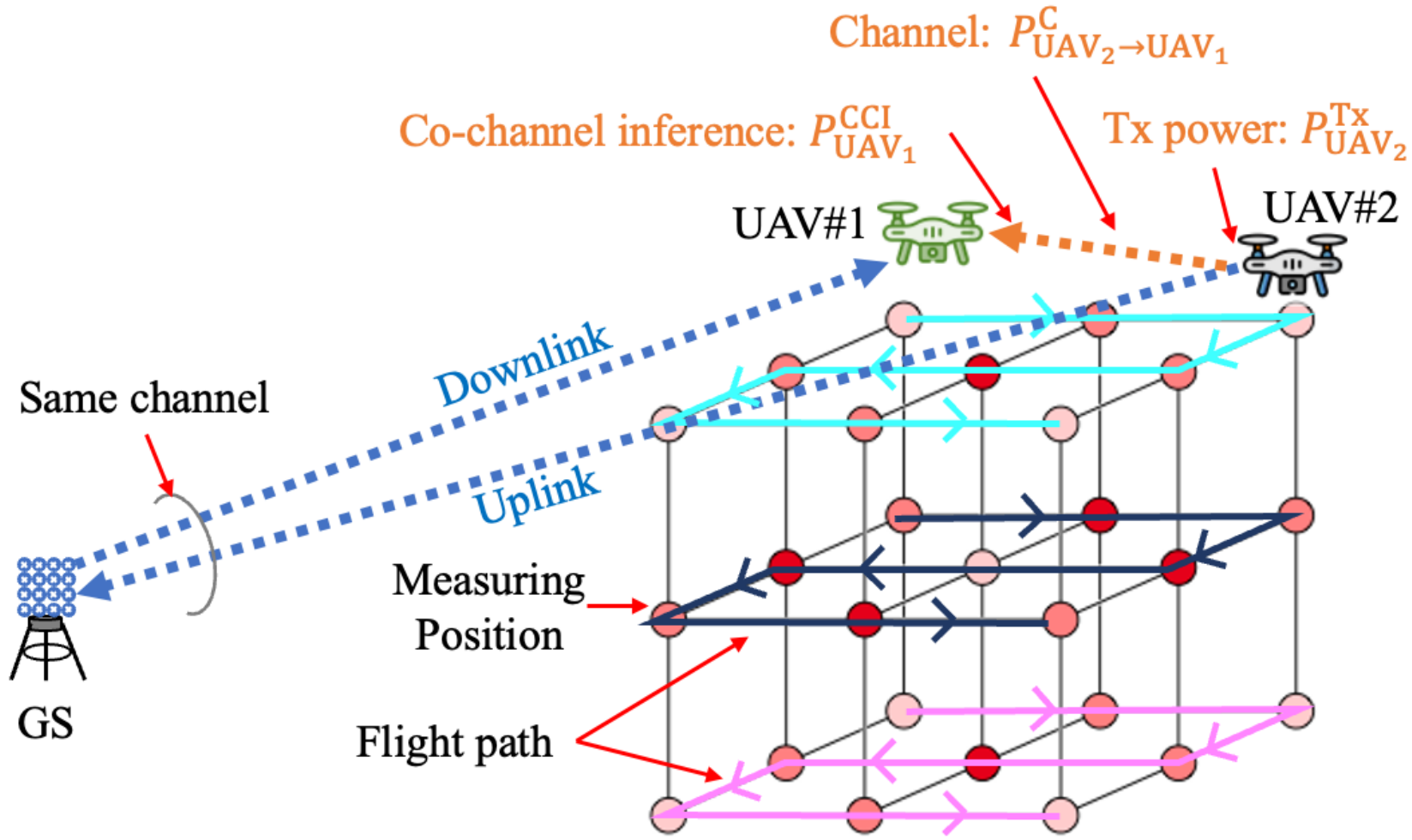}
    \caption{Measurement grid for co-channel interference between UAVs}
    \label{fig:metric}
\end{figure}

To evaluate whether the proposed system can effectively eliminate CCI and enable MU-IBFD, we measure the co-channel interference power $P^{\mathrm{CCI}}_1$ between UAVs as a function of their positions $(\boldsymbol r_1, \boldsymbol r_2)$. Because the interference is symmetric between the two UAVs, we only measure the interference from UAV\#2 uplink to UAV\#1 downlink using the same channel. The grid of measurement points around UAV\#2 and the relative geometry of the GS and two UAVs are illustrated in Fig.~\ref{fig:metric}.

\begin{figure}
    \centering
    \includegraphics[width=.48\textwidth]{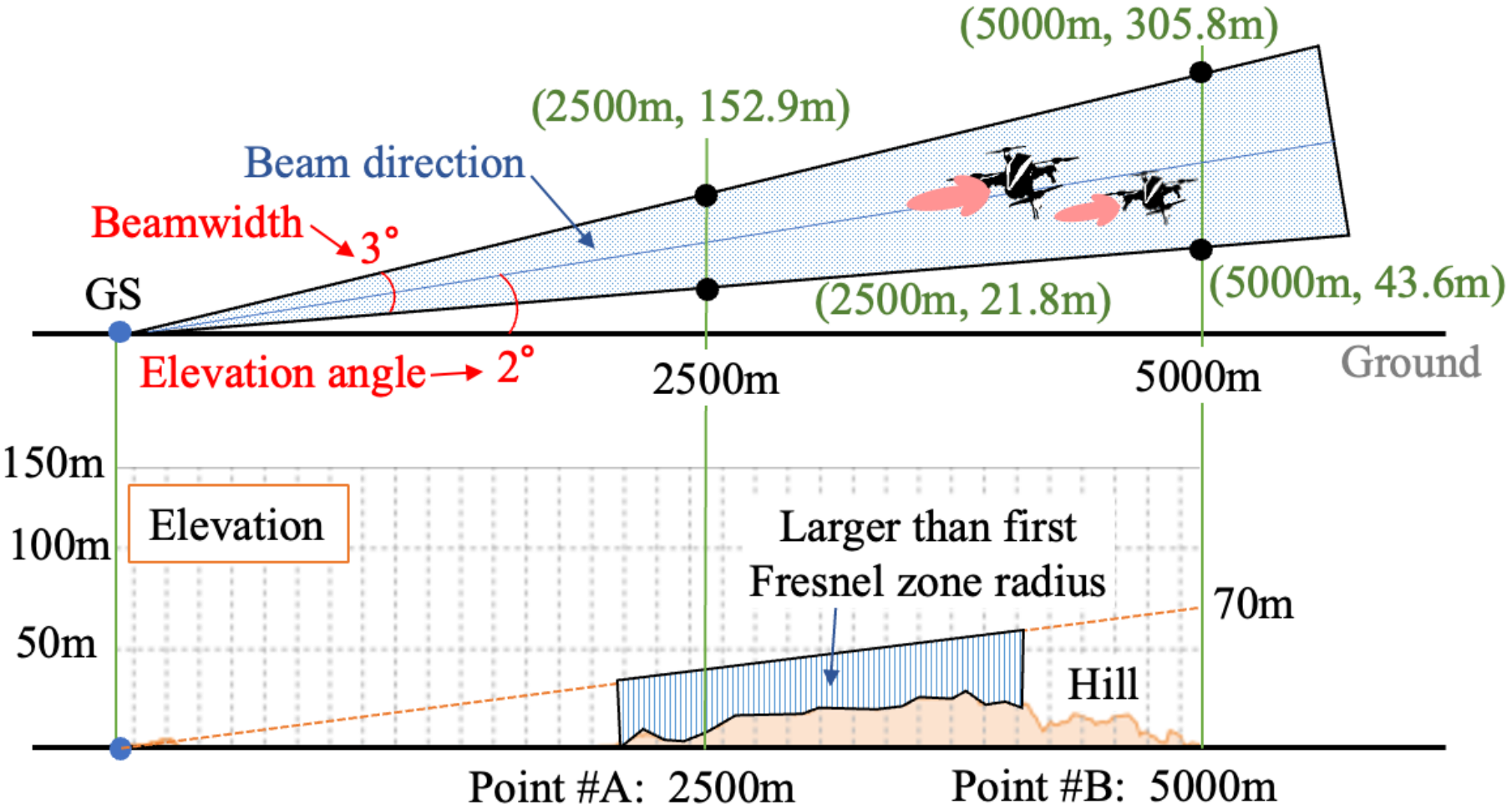}
    \caption{Side view of the fixed GS antenna setup and terrain profile between the GS and UAVs}
    \label{fig:fixgsantenna}
\end{figure}

On the coast of Minamisoma, Fukushima, Japan, experiments were conducted, and UAVs were at distances of around 2.5 km and 5 km away from GS and heights of 70m$\sim$100m.
A fixed GS antenna setup was employed instead of dynamic beam tracking, because UAVs were always within HPBW coverage of GS antenna in the experiments, due to the long GS-UAV distance. The side view of GS antenna setup and the vertical coverage are shown in Fig.~\ref{fig:fixgsantenna} for illustration. The terrain elevations between GS and UAV are also illustrated in the figure, in which it can be confirmed that the distance between obstacles (hills) and the GS-UAV straight-line path is always larger than the radius of the first Fresnel zone, so that a clear LOS always exists in the experiments.

\begin{figure}
    \centering
    \includegraphics[width=.48\textwidth]{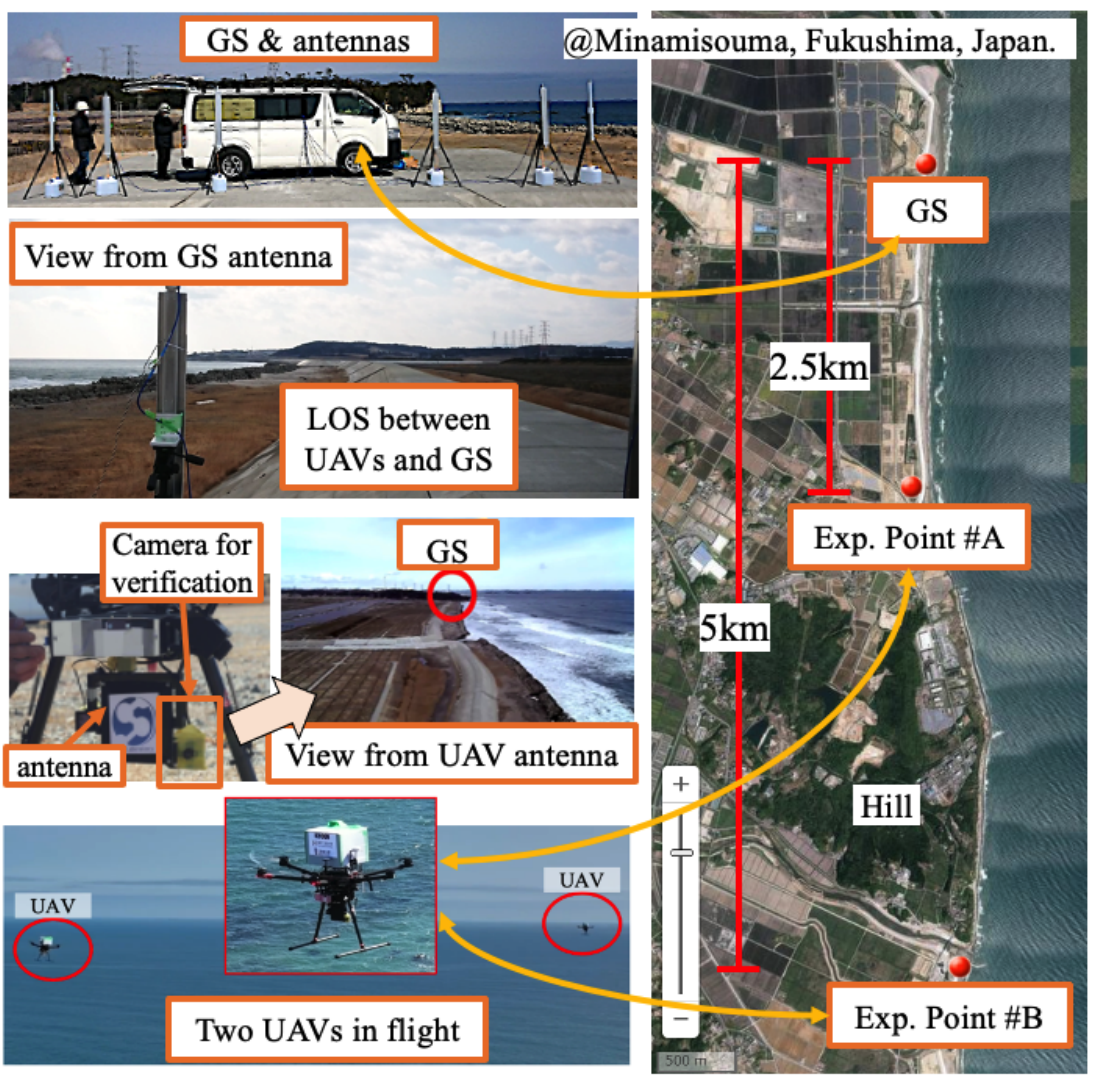}
    \caption{Experiment setup and environment in Minamisoma, Fukushima, Japan}
    \label{fig:expenv}
\end{figure}

Fig.~\ref{fig:expenv} depicts the experiment setup and environment. The GS with slot antennas was set up on a seawall built after the earthquake and tsunami on March 11, 2011 in Japan. LOS between UAVs and GS is shown by a photo of the view from GS antennas. A light-weight camera for afterwards verification of steering control is attached to the side of UAV antenna, and a photo taken by this camera is also shown. UAV antennas were always directed towards the GS by the mechanical rotator.

\begin{figure}
    \centering
    \includegraphics[width=.48\textwidth]{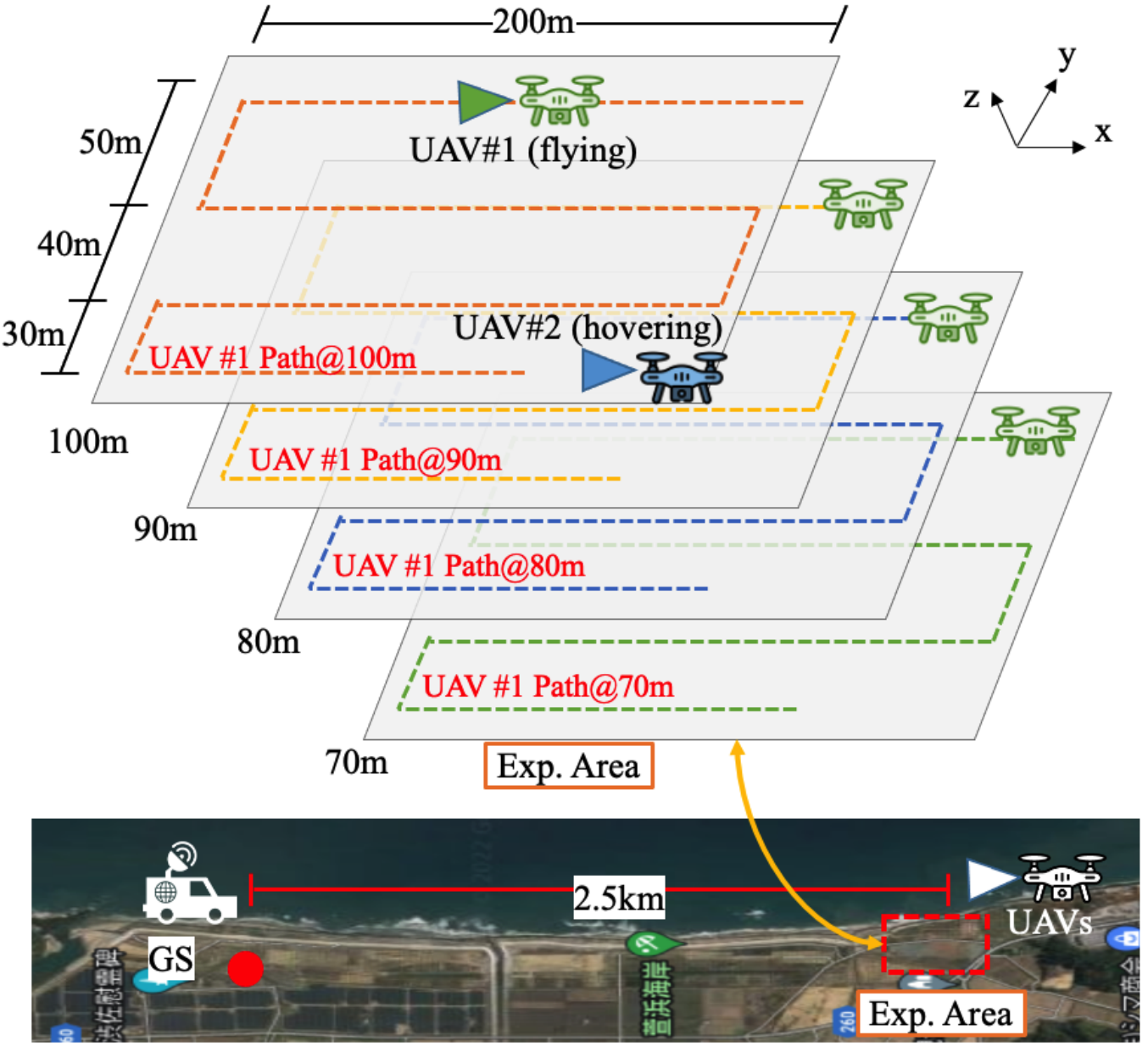}
    \caption{UAV\#1 flight paths for co-channel interference and signal-of-interest measurements at point \#A}
    \label{fig:expflight}
\end{figure}

At experiment point \#A, approximately 2.5 km away from GS, a quantitative experiment was conducted to evaluate the downlink performance. (Please note that experiment in this paper focused more on evaluating the downlink performance using the proposed MU-IBFD architecture.) The power of co-channel interference and signal of interest were measured when UAVs were in different relative positions. In the experiment shown in Fig.~\ref{fig:expflight}, UAV\#2 hovered at a fixed position at a height of 100m and transmitted 4K video to GS via uplink, while UAV\#1 flew near UAV\#2. The co-channel interference from UAV\#2 uplink to UAV\#1 downlink was measured and logged in UAV\#1. During co-channel interference measurements, GS turned off transmission to UAV\#1 via downlink, which uses the same channel as UAV\#2 uplink, so that the received signal at UAV\#1 was only co-channel interference from UAV\#2 uplink. Following co-channel interference measurements, UAV\#2 uplink was deactivated and the UAV\#1 downlink was activated, allowing the measurement of power of the signal of interest from GS and the calculation of UAV\#1 SINRs at various positions. The flying path of UAV\#1 at heights of 100m, 90m, 80m, and 70m are illustrated in Fig.~\ref{fig:expflight}. Since UAV\#2 hovered at a fixed height of 100\,m, these flight layers correspond to inter-UAV altitude differences of 0--30\,m, so that the vertical dimension of the ROR model is directly covered by the measurements. The received power was logged at 1\,Hz. For the interference map, UAV\#1 flew slowly at about 1\,m/s and repeated each altitude three times, while a separate pass at a typical cruise speed of about 5\,m/s was also recorded to capture the in-flight variation. The UAV positions were obtained by RTK-GPS with centimeter-level accuracy, so the position uncertainty of the measurement points is negligible. Since measurements were discrete, to make the results easier to follow, measurements were interpolated using Gaussian process regression for clarity. Given that two UAVs use antennas with identical patterns and that the co-channel interference between them is symmetric, only the interference from UAV\#2 to UAV\#1 was measured.

Near experiment point \#B (around 5 km away from GS), a proof-of-concept (PoC) demonstration was conducted to verify the feasibility of the proposed system. Two UAVs with their antennas oriented towards GS flew to a height of 100 m with a mutual distance of 50 m and performed MU-IBFD communication. 
(Their spatial configuration can be considered an isosceles triangle, with GS located at the apex, and two UAVs positioned at the vertices of the base.)
A photo, in which two UAVs were flying, is given at the bottom of Fig.~\ref{fig:expenv}.

It is noteworthy that the deliberate divergence in location selection is based on the following considerations: 
1) the point \#A with 2.5 km distance was chosen for interference testing, because this location offers a large open space characterized by a broader field of view and more stable airflow conditions, which is conducive to interference measurements with accurate UAV relative positions. It is important to note that in the proposed MU-IBFD architecture, co-channel interference is only dependent on relative positions of two UAVs and remains independent of distance between UAVs and GS.
2) the choice of the point \#B with 5 km distance for communication performance PoC was motivated by the necessity to validate system feasibility over an extended distance.

\section{EXPERIMENT RESULTS}
This section presents and discusses the results of co-channel interference measurement experiment and PoC demonstration of the proposed MU-IBFD system, which are conducted at points \#A and \#B, respectively, as explained above.

\subsection{Interference Measurement }

\begin{figure}
    \centering
    \includegraphics[width=.48\textwidth]{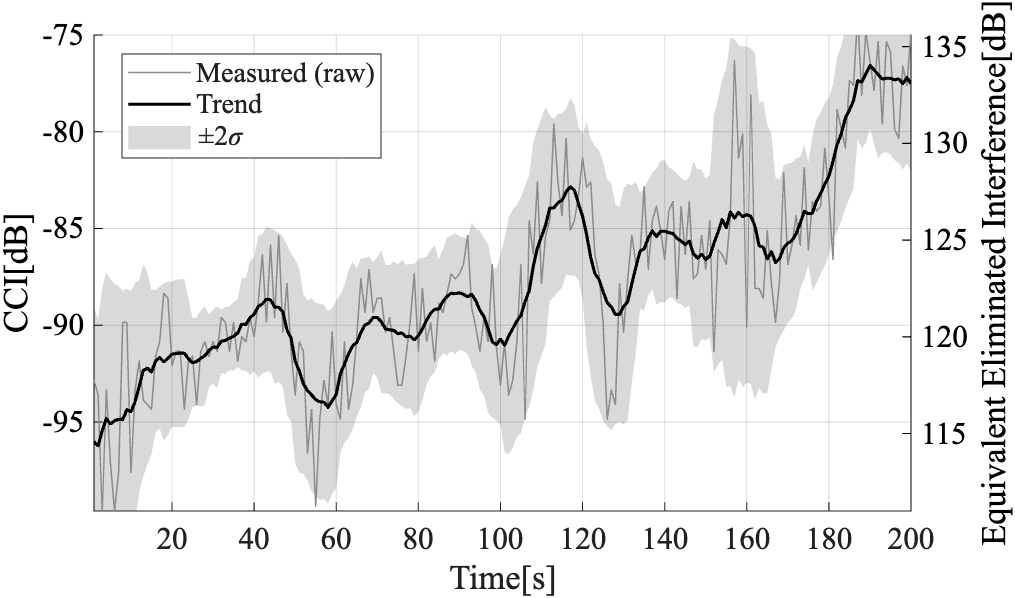}
    \caption{Measured co-channel interference of UAV\#1 at height of 100m (raw 1-Hz samples, moving-average trend, and $\pm2\sigma$ fluctuation band)}
    \label{fig:cciexample}
\end{figure}

\begin{figure}
    \centering
    \includegraphics[width=.48\textwidth]{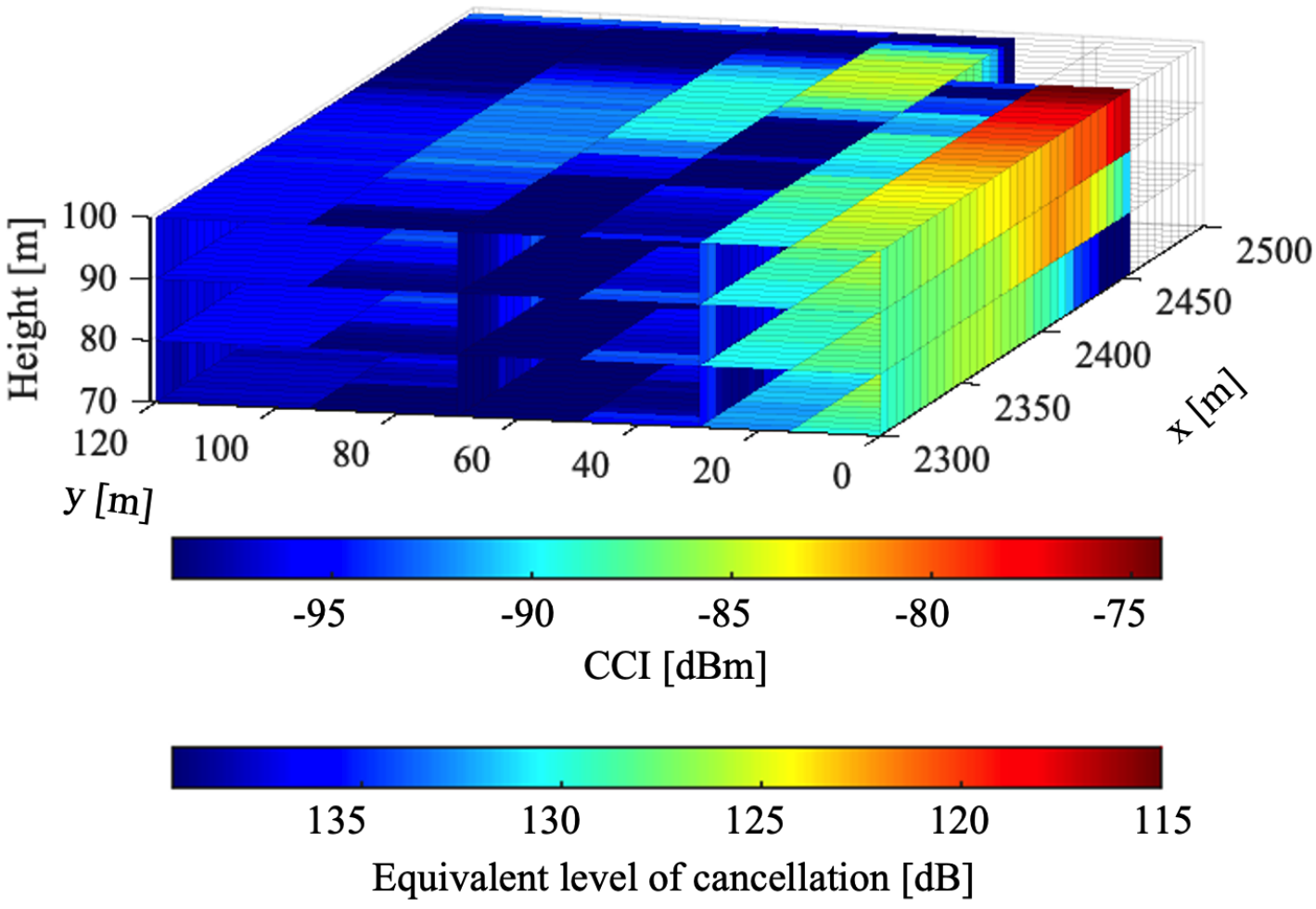}
    \caption{Interpolated co-channel interference at UAV\#1 and the corresponding equivalent cancellation level as functions of the UAV positions}
    \label{fig:results}
\end{figure}

Measurements of co-channel interference power and expected signal power were conducted multiple times at various heights of UAV\#1. It is noted that the measured co-channel interference also includes receiver noise.
The measured CCI is equivalent to the remaining interference after SIC in conventional IBFD. Fig.~\ref{fig:cciexample} shows an example of co-channel interference measured by UAV\#1 along the path in Fig.~\ref{fig:expflight} at a height of 100\,m, recorded during the cruise-speed pass. The figure shows the raw 1-Hz samples together with their moving-average trend and the $\pm2\sigma$ band of the local fluctuation. The fluctuation standard deviation is about $2.3$\,dB (the residual also contains trend-tracking error). For the slowly flown and thrice-repeated passes used for the interference map, each location aggregates several tens of samples, so the standard error of the values plotted in Fig.~\ref{fig:results} is a few tenths of a dB, far below the shadowing margin. In addition, the equivalent level of cancellation in conventional IBFD systems, with the equivalent isotropically radiated power (EIRP) of 36\,dBm, is also shown in Fig.~\ref{fig:cciexample} (right axis). It indicates that, to achieve the same downlink performance under this experimental condition, more than 110\,dB of self-interference needs to be eliminated by isolation and SIC, which is difficult to afford and complete in UAV-based communication systems.

For ease of understanding, all measured co-channel interference data are interpolated and summarized in Fig.~\ref{fig:results} (upper color bar). UAV\#2 is located at $(2500, 0, 100)$, and the GS at $(0, 0, 0)$. For safety reasons, as advised by UAV operators, measurements were not taken right above UAV\#2, leaving that area blank in the figure. The plot clearly shows that the CCI at UAV\#1 varies strongly with the relative positions of the two UAVs: a localized region of high interference is observed around UAV\#2, while most of the measurement area experiences much lower CCI. The equivalent level of cancellation, i.e., $C^{\rm eq}$, is shown in Fig.~\ref{fig:results} (lower color bar).

\subsection{ROR}

\begin{figure}
\centering
\subfloat[Calculated ROR (yellow points indicate the non-ROR region where the CCI constraint is violated)]{\includegraphics[width=.48\textwidth]{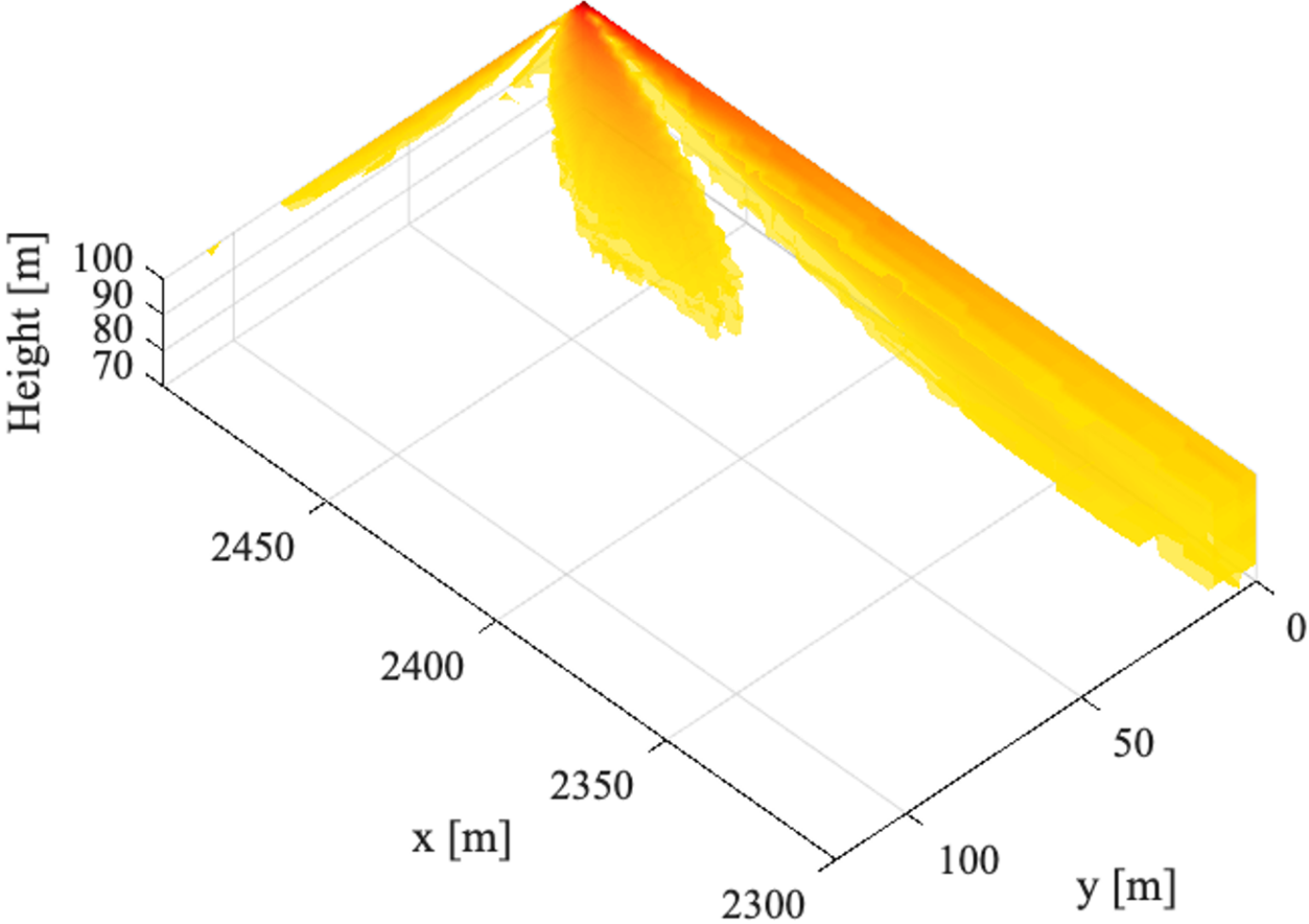}
\label{fig:ror}}
\hfill
\subfloat[Measured positions with $P^{\text{CCI}}_1 > I_{\text{th}}$]{\includegraphics[width=.48\textwidth]{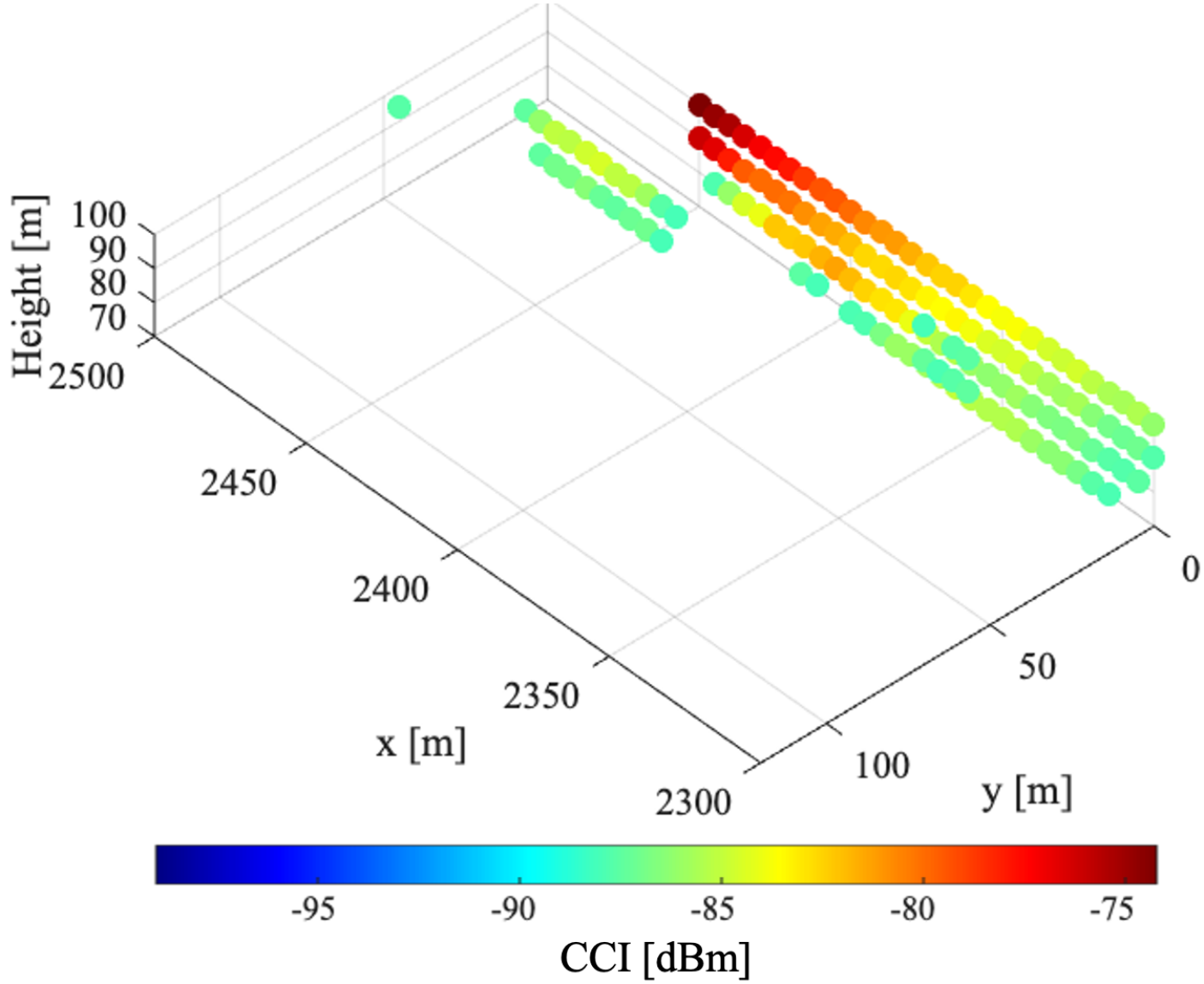}
\label{fig:cci}}
\hfill
\caption{Calculated reliable operating region and measured high-interference positions at point \#A.}
\label{fig:rorr}
\end{figure}

Given the positions of two UAVs, i.e., $(\boldsymbol r_1,\boldsymbol r_2)$, the ROR at point \#A can be numerically calculated from the condition in~\eqref{eq:safe_theta}. As an example, the target SINR was set to $\gamma_{\min}=10$ dB and a shadowing margin of $M_{\rm sh}=3$ dB was adopted.
The resulting ROR occupies most region around UAV\#2 considered in the experiment, and only a small subset of positions violates the CCI constraint. For better visibility, Fig.~\ref{fig:ror} therefore plots the complement of the ROR, i.e., the positions where the minimum SINR cannot be satisfied. The calculation covers the quarter space on one side of and below UAV\#2, and the remaining space follows from the symmetry of the antenna patterns. The colored areas in Fig.~\ref{fig:ror} represent the non-ROR region, whereas all the remaining space belongs to the ROR. It can be seen that the non-ROR region is confined to a narrow area very close to UAV\#2 and aligned with the main-lobe direction of antenna. The small patch on the side is caused by the sidelobes. This illustrates that, considering the typical usage of UAVs, for the considered system parameters, the ROR is extremely large.

The size of the non-ROR region can be further quantified using the exclusion distance $d_{\rm excl}$ introduced in Section~II-D. According to the calculated data in Fig.~\ref{fig:ror}, the non-ROR region consists of two parts. The first part is a narrow corridor along the main-lobe direction (i.e., the UAV\#2-to-GS direction). This corridor is confined within a cone with a half-angle of only about $7.5^{\circ}$, which occupies less than $0.5\%$ of the full solid angle. The second part is the small patch caused by the sidelobes, which lies within about $110$\,m from UAV\#2. In fact, in all directions more than $15^{\circ}$ away from the main-lobe axis, the exclusion distance is smaller than about $120$\,m. The corridor in the first part extends along the uplink main beam beyond the surveyed range; however, in the vast majority of operational geometries, such a relative position does not occur in the star topology, because a UAV staying in this corridor would also block the LOS between its companion UAV and the GS, and UAVs served by the same GS typically operate side by side, as in the PoC configuration in Section~III. Moreover, even if a mission does require placing a UAV in this direction, the position can simply be assigned to a UAV belonging to a different UAV pair, which reuses different channels and thus causes no CCI to the pair in question. In summary, IBFD operation is guaranteed as long as each UAV stays outside the keep-out zone around its companion UAV, which is narrow in solid angle and bounded in all off-axis directions.

To validate this geometry-aware model, the measured CCI values were compared with the interference threshold $I_{\text{th}}$, and the measurement points where $P^{\text{CCI}}_1 > I_{\text{th}}$ are plotted in Fig.~\ref{fig:cci}. These high-interference points cluster exactly in the non-ROR region. In contrast, all measured positions falling inside the ROR satisfy the minimum SINR constraint. The agreement between the calculated non-ROR region in Fig.~\ref{fig:ror} and the measured high-CCI positions in Fig.~\ref{fig:cci} confirms that the proposed ROR model accurately captures the interference behavior of the real system.

These results show that the ROR provides a physically meaningful and practically useful design tool: once the system parameters (antenna patterns, transmit power, SINR target, and shadowing margin) are fixed, the flight controller can restrict UAV operations to the large ROR in order to guarantee downlink reliability, while avoiding only a very small non-ROR region around each UAV. In other words, the MU-IBFD architecture can achieve near-ideal full-duplex performance over most of the 3D airspace by simple geometry-aware position management, without requiring on-board self-interference cancelers.

\subsection{Spectrum Efficiency}

\begin{figure}
    \centering
    \subfloat[Downlink capacity improvement compared to TDD and omni-directional antenna scheme]{\includegraphics[width=.48\textwidth]{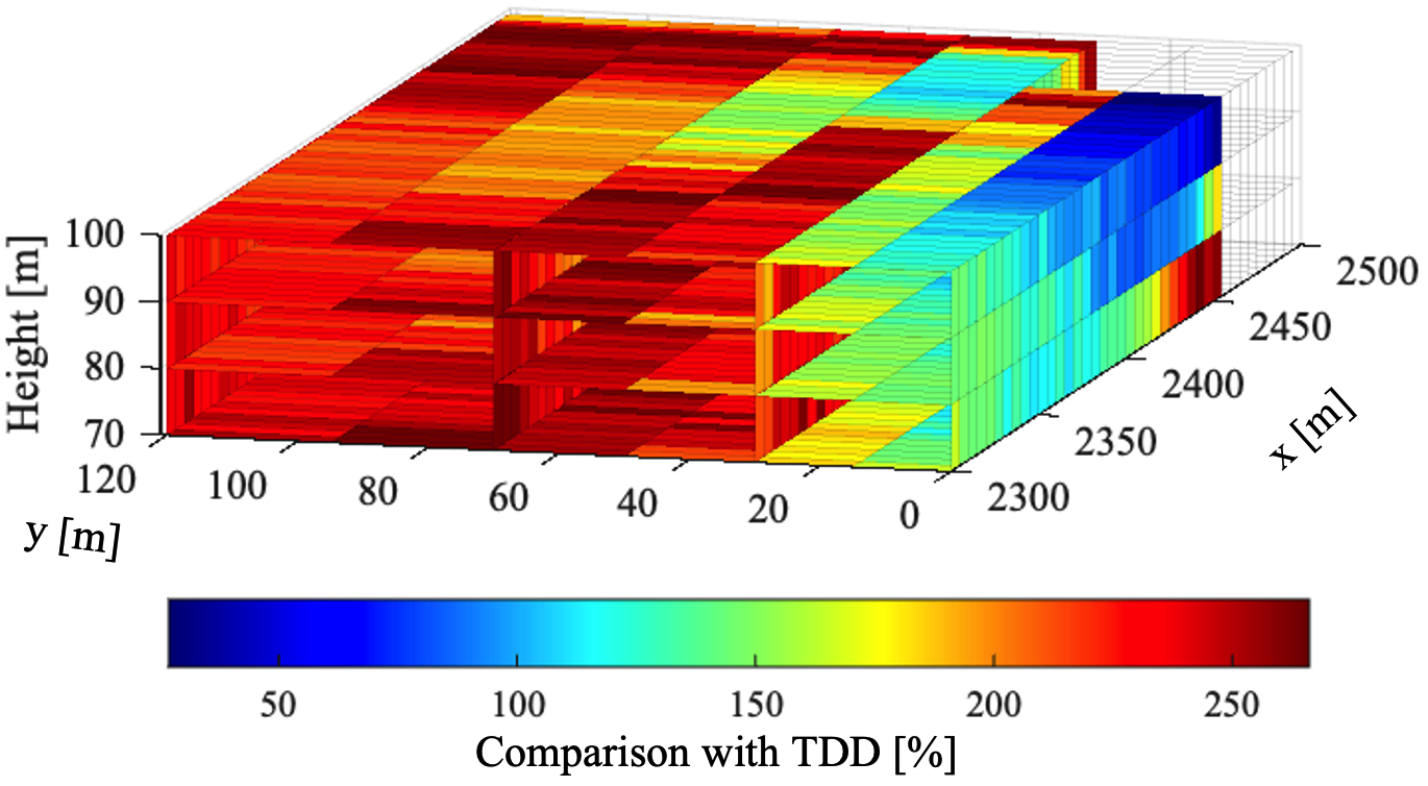}
    \label{fig:capacity}}
    \hfill
    \subfloat[Theoretical comparison of radio resource efficiencies with extended baselines]{\includegraphics[width=.48\textwidth]{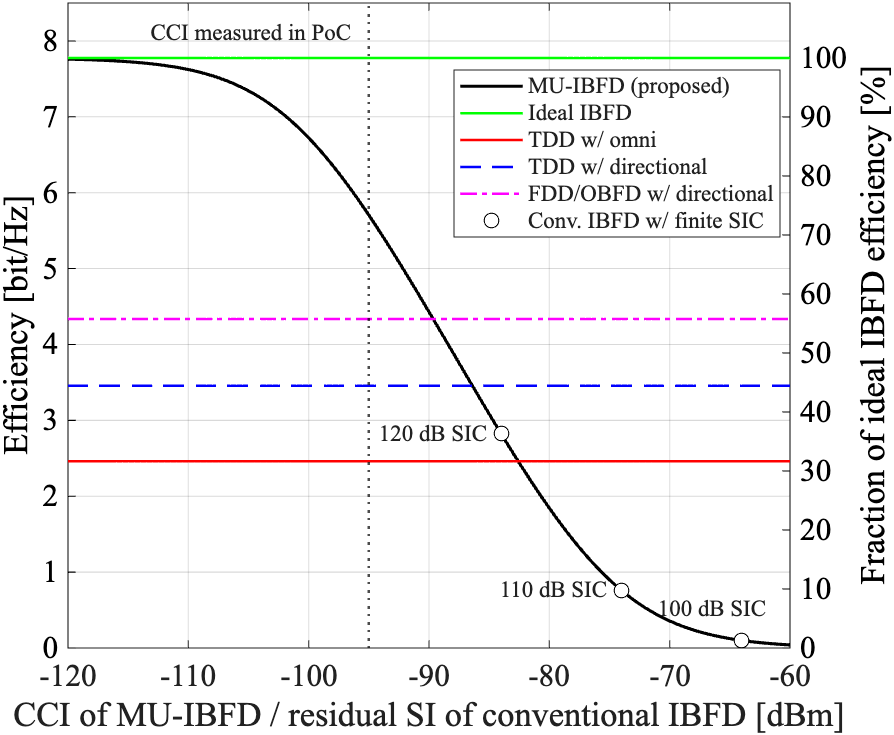}
    \label{fig:ccicap}}
    \hfill
    \caption{Radio resource efficiency of the proposed MU-IBFD system compared with TDD/FDD schemes, ideal IBFD, and conventional IBFD with finite SIC}
    \label{fig:eff}
\end{figure}

The channel capacity of the UAV\#1 downlink can be calculated from the measured co-channel interference power and signal-of-interest power in the experiment. To show the performance improvement, the measured capacity of the proposed system (with a 10\% guard band) is compared with the capacity of a conventional TDD/omni-directional antenna scheme with a 20\% guard interval, 36 dBm EIRP, and an ideal omni-directional antenna. 
The gain of MU-IBFD compared to the conventional scheme in downlink at different positions is calculated as
\begin{eqnarray}
\label{eq:comp}
R=\frac{C_{\text{MU-IBFD}}}{C_{\text{TDD}}}\times100\%
\end{eqnarray}
where $C_{\text{MU-IBFD}}$ and $C_{\text{TDD}}$ are the channel capacities of the MU-IBFD and TDD systems, respectively.
The results in Fig.~\ref{fig:capacity} show that the MU-IBFD system can obtain a significantly better downlink capacity than the conventional TDD system, even though the latter uses a higher Tx power, except in the blue area in Fig.~\ref{fig:capacity} near the main-lobe direction of UAV\#2. 
Please note that the measurements of co-channel interference and the corresponding calculation of capacity in this experiment are limited by the noise floor of the UAV transceiver.
To compare the proposed system with a broader set of baselines under an identical link budget, Fig.~\ref{fig:ccicap} presents a theoretical comparison of the radio resource efficiencies as functions of the interference power: conventional TDD with an omni-directional antenna and 36\,dBm EIRP (the same baseline as in Fig.~\ref{fig:capacity}), TDD and FDD/OBFD with the same directional antennas as the proposed system, ideal IBFD with perfect SIC, and conventional IBFD with finite SIC. Since the residual self-interference after SIC plays the same role in \eqref{eq:sinr1} as the CCI, the black curve can be read in two ways: for the proposed MU-IBFD, the horizontal axis is the CCI determined by the UAV geometry, whereas for a conventional IBFD transceiver, it is the residual self-interference $P_{\mathrm{SI,0}}/C$ in \eqref{eq:Ceq}, with the markers indicating the operating points for $C=100$--$120$\,dB (with $P_{\mathrm{SI,0}}=36$\,dBm, as in Fig.~\ref{fig:cciexample}). The right y-axis normalizes all efficiencies to the ideal-IBFD upper bound, and the vertical dotted line marks the CCI of about $-95$\,dBm measured in the PoC (see Section~IV-D), where about $73\%$ of the ideal bound is achieved.

Three trends can be observed from Fig.~\ref{fig:ccicap}. First, upgrading the antennas alone does not close the gap: with the same directional antennas, the TDD and FDD/OBFD baselines recover only a part of the loss, and the major share of the gain of the proposed system comes from the full-duplex channel reuse enabled by the architecture. Second, the efficiency of conventional IBFD is dictated by its SIC capability and collapses once the residual self-interference approaches the signal level; reaching the operating point of the prototype would require more than $130$\,dB of cancellation, far beyond what is affordable on UAVs. Third, and most importantly, the operating point of a conventional IBFD transceiver on this curve is fixed by its hardware, whereas in the proposed system the CCI is controlled by the UAV geometry: the system can thus always be operated on the favorable left side of the curve and approaches the ideal IBFD bound as the CCI decreases. It is noted that a direct numerical comparison with the UAV prototypes reported in the literature is infeasible due to the different frequency bands, hardware, and scenarios; the baselines are therefore evaluated under the identical link budget and measured data of the prototype.

\subsection{PoC Demonstration}

\begin{figure}
    \centering
    \subfloat[Video from two UAVs at GS]{\includegraphics[width=.48\textwidth]{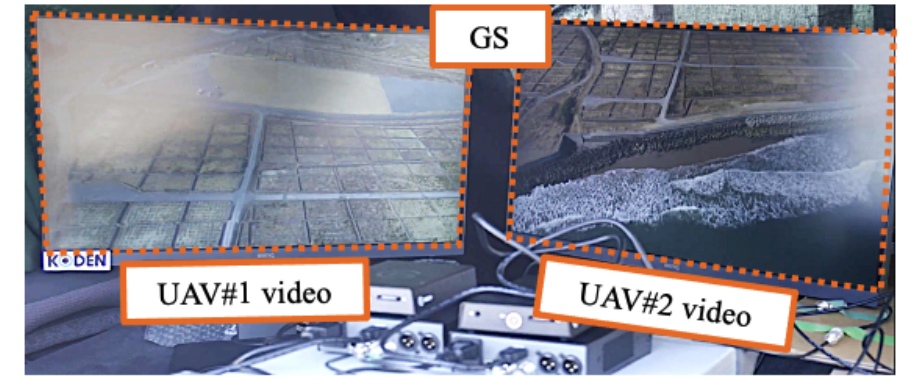}
    \label{fig:twovideo}}
    \hfill
    \subfloat[Uplink and downlink performance of one UAV]{\includegraphics[width=.48\textwidth]{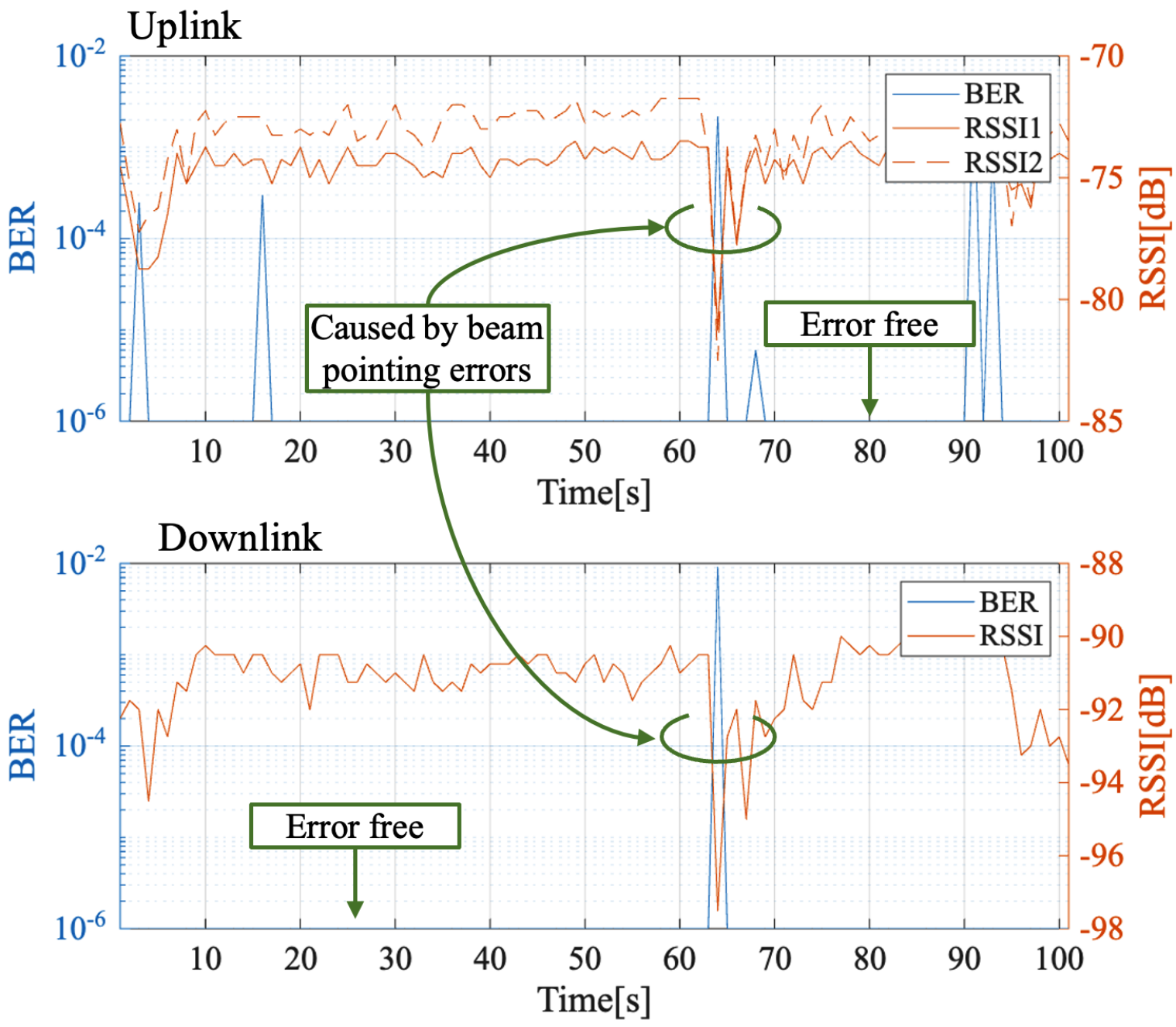}
    \label{fig:updownrssi}}
    \hfill
    \caption{PoC demonstration of 5\,km dual 4K video transmission and link performance of one UAV}
    \label{fig:poc}
\end{figure}

Fig.~\ref{fig:twovideo} depicts two monitors displaying 4K videos received at GS, showing that two streams of 4K video were successfully transmitted from UAVs 5 km away to GS, when two UAVs flew at a height of 100 m with a mutual distance of around 50 m.
The received signal strength indicators (RSSIs) and BERs were also measured and logged by GS and UAVs. As an example, these metrics of both uplink and downlink in one UAV are shown in Fig.~\ref{fig:updownrssi} for 100 seconds of transmission. The BER of $10^{-6}$ in the figure is considered an error-free transmission. This demonstration verifies that MU-IBFD communications were established by the prototype system.

As explained earlier, the diminished RSSI observed in downlink, depicted in Fig.~\ref{fig:updownrssi}, stems from the deliberate design of downlink Tx power in accordance with the requirements of the application. It is noteworthy that higher RSSI can be achieved by increasing the Tx power, if necessitated by the application requirements. 
The assessment of co-channel interference, as presented in the previous sub-section and illustrated in Fig.\ref{fig:cciexample} and Fig.\ref{fig:results}, reveals a co-channel interference level of approximately -95 dBm in this PoC, while in the context of conventional IBFD systems, it is necessary to achieve a 135 dB of equivalent interference cancellation for comparable performance.

In the experiment, occasional RSSI drops accompanied by short error bursts were observed, as illustrated in Fig.~\ref{fig:updownrssi}. The drops occur in the uplink and in the downlink at the same instants, and since the two directions share the same UAV antenna while the inter-UAV geometry and hence the CCI remained unchanged, they can be attributed to the beam pointing rather than to the propagation or the interference. Quadrotor UAVs, including the DJI M600Pro used here, frequently adjust the attitude in response to wind and turbulence, and a gust-induced attitude transient can momentarily exceed the tracking bandwidth of the mechanical rotator, which produces exactly such simultaneous short fades on both directions. These events are rare, occupying only a few seconds in total within the 100-s record, and the transmission stayed error-free elsewhere. To address this, future plans include incorporating a stabilizer and 3-dimensional rotator for improved antenna steering control and stabilization.

\section{Practical Considerations}

The field trials in Section~IV were conducted in open airspace with LOS conditions and without co-existing systems in the same band, using prototype-grade hardware. This section discusses the behavior of the proposed framework beyond these conditions, as well as practical energy and form-factor considerations.

\textit{1) Non-LOS conditions:} For the UAV-to-UAV interference link, blockage only attenuates the CCI, so the LOS-based model in Section~II is a worst-case characterization, and the ROR only becomes more conservative. For the GS-to-UAV desired link, obstructions reduce $P_1^{\rm S}$ and thereby the threshold $I_{\rm th}$ in \eqref{eq:Ith}. Since $P_1^{\rm S}$ is routinely measured by the UAVs and reported to the GS (Section~II-D), the GS-side flight controller, which maintains the positions of all UAVs in the star topology, can update the contracted ROR and coordinate the UAV positions accordingly, optionally aided by prior terrain information. The framework thus adapts through routine link reports and centralized control, without relying on a propagation model of the environment; for severely disturbed environments such as disaster areas \cite{refB}, the probabilistic ROR extension discussed in Section~II can be applied within the same framework.

\textit{2) External interference:} Interference from terrestrial systems sharing the same band, or from UAVs outside the system, is mitigated in three ways: the directional antennas at both ends spatially filter the signals arriving from side- and back-lobe directions, and in particular, the narrow elevation beam of the GS antenna suppresses the waves from the ground; residual external interference enters the interference-plus-noise term and contracts $I_{\rm th}$, which is handled by the same reporting and centralized control mechanism as above; and the coexistence with terrestrial systems, e.g., a WLAN sharing the same 5.7\,GHz band, has been preliminarily studied in our earlier work \cite{share} by spectrum monitoring and pre-planned flyable areas. A general treatment of multi-system spectrum sharing is left for future work.

\textit{3) Energy efficiency and hardware compactness:} The communication subsystem burdens a UAV in two ways, namely the power drawn by the transceiver and the payload weight coupled to the kilowatt-level propulsion power of the multirotor platform \cite{dji}. The proposed architecture is economical in both. The link budget is carried by the combined passive antenna gains of about 34\,dB rather than by the transmit power, so the UAV PA outputs only 17\,dBm, whereas the omni-directional baseline must generate its entire 36\,dBm EIRP by the PA alone, and the full-duplex reuse further reduces the energy per delivered bit in high-rate and long-duration missions. Meanwhile, removing the SIC chain saves not only its power but also its weight. Regarding the form factor, the bulk of the current prototype stems from the implementation, such as the FPGA prototyping boards and the general-purpose rotator, rather than from the architecture, and has been largely reduced in the newer compact transceiver. Replacing the mechanical rotator with an electronically steered array is a natural next step, which also removes the moving parts and improves the pointing robustness discussed in Section~II-D.

\section{CONCLUSION}

This study presents a field experiment using prototype hardware to verify the feasibility of the proposed geometry-aware MU-IBFD system. The results confirm that the proposed architecture can effectively reduce co-channel interference, which is equivalent to the self-interference in conventional IBFD systems. The SINR and capacity are greatly influenced by the relative positions of the UAVs, indicating that optimal positioning can significantly boost system performance. The results show that the MU-IBFD system outperforms the traditional TDD scheme. Future works include extending the system to multi-hop systems, and joint optimization of flight path planning and communication resource management.

\begin{IEEEbiography}
[{\includegraphics[width=1in,height=1.25in,clip,keepaspectratio]{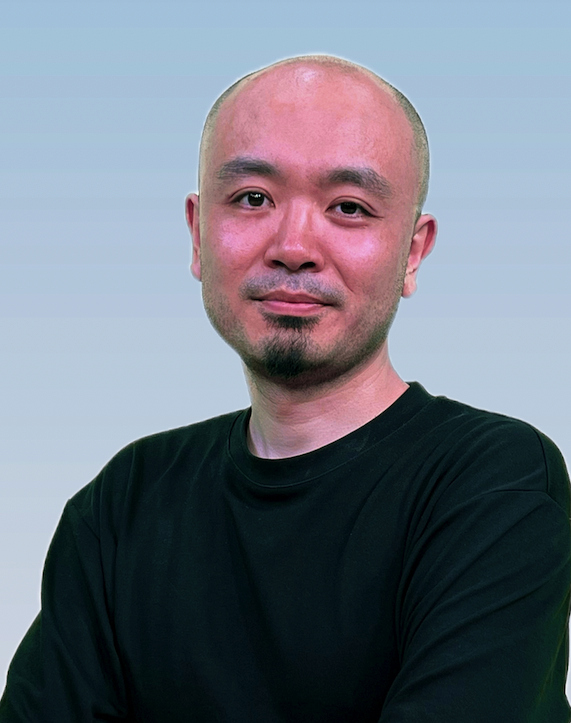}}]
{Tao Yu}{\space}(Member, IEEE) received a M.E. degree from the Communication University of China in 2010, and a Dr.Eng. degree from the Tokyo Institute of Technology (now Institute of Science Tokyo, Science Tokyo) in 2017. He worked as a researcher from 2017 to 2022 in Dept. Electr. Electron. Eng. at Tokyo Tech. Since 2022, he has been a specially appointed associate professor at the Academy of Super Smart Society, Science Tokyo. His research interests include smart mobility, autonomous driving, digital twin, V2X, UAV communication, mmWave, sensor networks, localization, antenna design, and building energy management. He is a member of IEEE and IEICE.
\end{IEEEbiography}%

\begin{IEEEbiography}[{\includegraphics[width=1in,height=1.25in,clip,keepaspectratio]{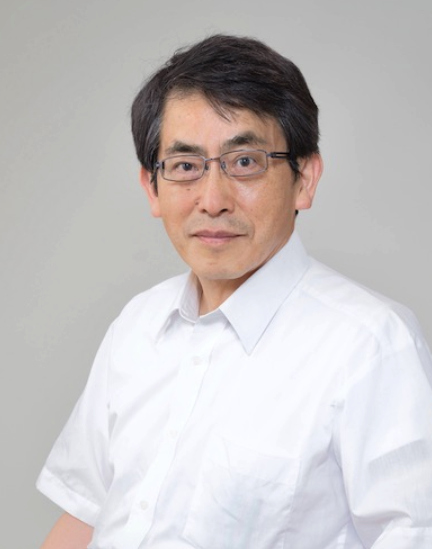}}]
{Kiyomichi Araki}{\space} received the B.S. degree in electrical engineering from Saitama University, in 1971, and the M.S. and Ph.D. degrees in physical electronics both from Tokyo Institute of Technology in 1973 and 1978 respectively. In 1973-1975, and 1978-1985, he was a Research Associate at Tokyo Institute of Technology, and in 1985-1995 he was an Associate Professor at Saitama University. In 1979-1980 and 1993-1994 he was a visiting research scholar at University of Texas, Austin and University of Illinois, Urbana, respectively. From 1995 to 2014 he was a Professor at Tokyo Institute of Technology. His research interests are in information security, coding theory, communication theory, ferrite devices, RF circuit theory, electromagnetic theory, software defined radio, array signal processing, UWB technologies, wireless channel modeling and so on. He is a honorary member of IEICE.
\end{IEEEbiography}

\begin{IEEEbiography}[{\includegraphics[width=1in,height=1.25in,clip,keepaspectratio]{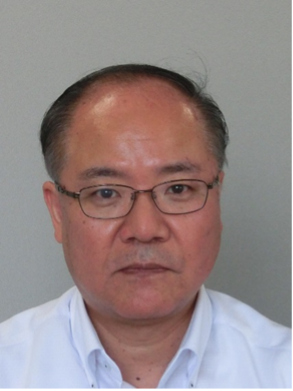}}]
{Tomohiro Mogi}{\space} received a B.S. degree from the Department of Precision Mechanics, Faculty of Science and Engineering, Chuo University in 1987, and in the same year he joined Yagi Antenna Co., Ltd. (now HYS Engineering Service Inc.), where he engaged in design and development of satellite communications, terrestrial broadcasting, and CATV-related equipment. He joint development department of Koden Electronics Co., Ltd. in July 2018, where he engaged in research and development of high-definition video real-time wireless transmission and now he serves as the Environmental Management Representative.
\end{IEEEbiography}

\begin{IEEEbiography}
[{\includegraphics[width=1in,height=1.25in,clip,keepaspectratio]{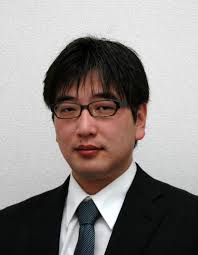}}]
{Yasushi Hada}{\space} received his Ph.D. in Engineering from the University of Tsukuba, Graduate School of Engineering, in 2003. He began his career at RIKEN (The Institute of Physical and Chemical Research) in the same year, and later joined the NICT (National Institute of Information and Communications Technology) in 2007. In 2011, he joined the Department of Mechanical Systems Engineering at Kogakuin University as an Associate Professor, a position he currently holds. His research focuses on communication systems for disaster response robots, autonomous robots, construction robots, intelligent environments, and service engineering. He is a member of several professional societies, including the Robotics Society of Japan (RSJ), the Information Processing Society of Japan (IPSJ), and the Japan Society of Mechanical Engineers (JSME).
\end{IEEEbiography}

\begin{IEEEbiography}[{\includegraphics[width=1in,height=1.25in,clip,keepaspectratio]{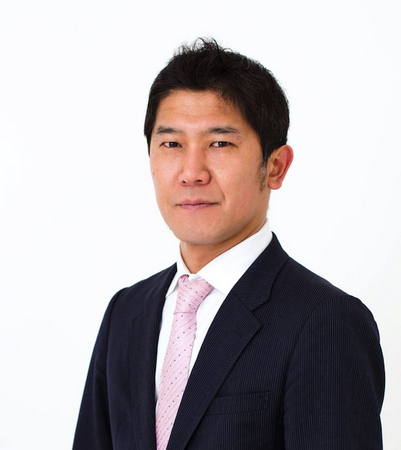}}]
{Kei Sakaguchi}{\space}(Senior Member, IEEE) is the Director of the Visionary Initiative (Innovative-Life Society), Dean of the Academy of Super Smart Society, and a Professor in the School of Engineering in the Institute of Science Tokyo. He received the M.E. degree in Information Processing from Tokyo Institute of Technology in 1998, and the Ph.D. degree in Electrical and Electronic Engineering from Tokyo Institute of Technology in 2006. He received the Outstanding Paper Awards from SDR Forum and IEICE in 2004 and 2005, respectively, and three Best Paper Awards from IEICE Communications Society in 2012, 2013, and 2015. He also received the Tutorial Paper Award from IEICE Communications Society in 2006. His current research interests are in 5G/6G cellular networks and digital twins for super smart society such as smart mobility, smart agriculture, smart ocean, and smart healthcare. He is a Fellow of IEICE and a Senior Member of IEEE.
\end{IEEEbiography}


\begin{thebibliography}{1}

\bibitem{1}
H. Shakhatreh, A.H. Sawalmeh, A. Al-Fuqaha, Z. Dou, E. Almaita, I. Khalil, N.S. Othman, A. Khreishah, and M. Guizani, ``Unmanned aerial vehicles (UAVs): A survey on civil applications and key research challenges'',
\newblock \emph{IEEE Access}, vol.7, pp.48572-48634, 2019.

\bibitem{2}
E. Zanelli, H. B\"{o}decker, ``Global Drone Market Report 2023-2030'', \newblock \emph{Drone Industry Insights}, July 2023.

\bibitem{4}
``Enhanced LTE support for aerial vehicles'', 
\newblock \emph{3GPP TR} 36.777, 2019.

\bibitem{5}
L. Bertizzolo, T. X. Tran, B. Amento, B. Balasubramanian, R. Jana,H. Purdy, Y. Zhou, and T. Melodia, ``Live and let Live: Flying UAVs Without Affecting Terrestrial UEs'', 
\newblock \emph{ACM HotMobile}, pp.21-26, 2020.

\bibitem{3}
S. Hayat, E. Yanmaz, and R. Muzaffar, ``Survey on unmanned aerial vehicle networks for civil applications: A communications viewpoint'', 
\newblock \emph{IEEE Commun. Surv. Tutor.}, vol.18.4, pp.2624-2661, 2016.

\bibitem{6}
Y. Takaku, T. Yu, Y. Kaieda, and K. Sakaguchi, ``Proof-of-Concept of Uncompressed 4K Video Transmission from Drone through mmWave'', 
\newblock \emph{IEEE CCNC}, pp.1-6, 2020.

\bibitem{7}
T. Yu, Y. Takaku, Y. Kaieda, and K. Sakaguchi, ``Design and PoC Implementation of mmWave-based Offloading-enabled UAV Surveillance System'', 
\newblock \emph{IEEE Open J. Veh. Technol.}, vol.2, pp.436-447, 2021.

\bibitem{8}
W. Khawaja, O. Ozdemir, and I. Guvenc, ``UAV air-to-ground channel characterization for mmWave systems'', 
\newblock \emph{IEEE VTC2017-Fall}, pp.1-5, 2017.

\bibitem{9}
K. E. Kolodziej, B. T. Perry and J. S. Herd, ``In-Band Full-Duplex Technology: Techniques and Systems Survey'', 
\newblock \emph{IEEE Trans. Microw. Theory Techn.}, vol.67, no.7, pp.3025-3041, Jul. 2019.

\bibitem{13}
S. Bojja Venkatakrishnan, E. A. Alwan and J. L. Volakis, ``Wideband RF Self-Interference Cancellation Circuit for Phased Array Simultaneous Transmit and Receive Systems'', 
\newblock \emph{IEEE Access}, vol.6, pp.3425-3432, 2018.

\bibitem{14}
E. Ahmed and A. M. Eltawil, ``All-Digital Self-Interference Cancellation Technique for Full-Duplex Systems'', 
\newblock \emph{IEEE Trans. Wirel. Commun.}, vol.14, no.7, pp.3519-3532, Jul. 2015.

\bibitem{15}
D. Korpi, L. Anttila and M. Valkama, ``Reference receiver based digital self-interference cancellation in MIMO full-duplex transceivers'', 
\newblock \emph{IEEE Globecom}, pp.1001-1007, 2014. 


\bibitem{10}
E. Everett, M. Duarte, C. Dick and A. Sabharwal, ``Empowering full-duplex wireless communication by exploiting directional diversity'', 
\newblock \emph{IEEE ASILOMAR}, pp.2002-2006, 2011.

\bibitem{11}
E. Everett, A. Sahai and A. Sabharwal, ``Passive Self-Interference Suppression for Full-Duplex Infrastructure Nodes'', 
\newblock \emph{IEEE Trans. Wireless Commun.}, vol.13, no.2, pp.680-694, Feb. 2014.

\bibitem{12}
G. Makar, N. Tran and T. Karacolak, ``A High-Isolation Monopole Array With Ring Hybrid Feeding Structure for In-Band Full-Duplex Systems'', 
\newblock \emph{IEEE Antennas Wirel. Propag. Lett.}, vol.16, pp.356-359, 2017.


\bibitem{16}
L.  Zhang,  Q.  Fan,  and  N.  Ansari, ``3-D  drone-base-station  placement with in-band full-duplex communications'', 
\newblock \emph{IEEE Commun. Lett.}, vol.22, no.9, pp.1902–1905, Sep. 2018.

\bibitem{17}
H. Wan, J. Wang, G. Ding, J. Chen, Y. Li, and Z. Han, ``Spectrum sharing planning for full-duplex UAV relay-ing systems with underlaid D2D communications'', 
\newblock \emph{IEEE J. Sel. AreasCommun.}, vol.36, no.9, pp.1986-1999, Sep. 2018.

\bibitem{18}
G. Liu, F. R. Yu and H. Ji, ``In-band full-duplex relaying: A survey research issues and challenges'', 
\newblock \emph{IEEE Commun. Surv. Tut.}, vol.17, no.2, pp.500-524, Jun 2015.

\bibitem{19}
L. Zhang and N. Ansari, ``A Framework for 5G Networks with In-Band Full-Duplex Enabled Drone-Mounted Base-Stations'', 
\newblock {IEEE Trans. Wirel. Commun.}, vol.26, no.5, pp.121-127, Oct. 2019.

\bibitem{c17}
T. Yu, Y. Takaku, Y. Kaieda and K. Sakaguchi, ``Design and PoC Implementation of Mmwave-Based Offloading-Enabled UAV Surveillance System'', 
\newblock \emph{IEEE Open J. Veh. Technol.}, vol. 2, pp. 436-447, 2021.

\bibitem{c15}
L. Sundqvist, ``Cellular controlled drone experiment: Evaluation of network requirements'', 
\newblock Master thesis, School Electr. Eng., Aalto University, Otaniemi, Espoo, Finland, 2015.

\bibitem{c14}
S.A. Hadiwardoyo, C.T. Calafate, J.C. Cano, Y. Ji, E. Hernández-Orallo, and P. Manzoni, ``Evaluating UAV-to-car communications performance: from testbed to simulation experiments'', 
\newblock \emph{IEEE CCNC}, Jan. 2019.



\bibitem{c16}
X. Lin, R. Wiren, S. Euler, A. Sadam, H.L. Määttänen, S. Muruganathan, S. Gao, Y.P. E. Wang, J. Kauppi, Z. Zou, and V. Yajnanarayana, ``Mobile network-connected drones: Field trials, simulations, and design insights'', 
\newblock \emph{IEEE Veh. Technol. Mag.}, vol. 14, no. 3, pp. 115–125, Sep. 2019.



\bibitem{c18}
Q. Song, Y. Zeng, J. Xu and S. Jin, ``A survey of prototype and experiment for UAV communications'', 
\newblock \emph{Sci. China Inf. Sci.}, vol. 64, 2021, art. no. 140301.

\bibitem{c12}
T. Yu, K. Araki and K. Sakaguchi, ``Full-Duplex Aerial Communication System for Multiple UAVs with Directional Antennas'', 
\newblock \emph{IEEE CCNC}, Jan. 2022.

\bibitem{c13}
T. Yu, S. Imada, K. Araki and K. Sakaguchi, ``Multi-UAV Full-Duplex Communication Systems for Joint Video Transmission and Flight Control'', 
\newblock \emph{IEEE CCWC}, Jan. 2021.

\bibitem{exp}
T. Yu, K. Kajiwara, K. Araki, and K. Sakaguchi, ``Experiment of Multi-UAV Full-Duplex System Equipped with Directional Antennas'', 
\newblock \emph{IEEE CCNC}, Jan. 2023.

\bibitem{dji}
DJI Matrice 600 Pro, [Online] Available:\url{https://www.dji.com/jp/matrice600-pro}

\bibitem{c19}
TSGR, ``TS 136 211 - V12.4.0 - LTE; Evolved Universal Terrestrial Radio Access (E-UTRA); Physical channels and modulation  (3GPP TS 36.211 version 12.4.0 Release 12)'',
\newblock \emph{vol. 0, 2015.}

\bibitem{refA}
O. S. Oubbati, J. Alotaibi, F. Alromithy, M. Atiquzzaman, and M. R. Altimania, ``A UAV-UGV Cooperative System: Patrolling and Energy Management for Urban Monitoring'',
\newblock \emph{IEEE Trans. Veh. Technol.}, 2025.

\bibitem{refC}
A. I. Ameur, O. S. Oubbati, A. Rachedi, A. Arishi, and M. Atiquzzaman, ``Intelligent UAV Caching and Energy Management in 6G Networks'',
\newblock \emph{IEEE Trans. Netw. Sci. Eng.}, 2025.

\bibitem{refB}
J. Alotaibi, O. S. Oubbati, M. Atiquzzaman, F. Alromithy, and M. R. Altimania, ``Optimizing disaster response with UAV-mounted RIS and HAP-enabled edge computing in 6G networks'',
\newblock \emph{J. Netw. Comput. Appl.}, vol. 241, art. no. 104213, 2025.

\bibitem{share}
T. Yu, K. Kajiwara, K. Araki, and K. Sakaguchi, ``Spectrum Sharing between Directional-Antenna-Equipped UAV System and Terrestrial Systems'',
\newblock \emph{IEEE CCWC}, Jan. 2022.




\end{thebibliography}
\end{document}